\documentclass[conference]{IEEEtran}
\IEEEoverridecommandlockouts
\usepackage{cite}
\usepackage{amsmath,amssymb,amsfonts}
\usepackage{algpseudocode}
\usepackage{algorithmicx}
\usepackage{algorithm}
\usepackage[caption=false]{subfig}
\usepackage{flushend}
\usepackage{multirow}
\usepackage{graphicx}
\usepackage{textcomp}
\usepackage{xcolor}
\def\BibTeX{{\rm B\kern-.05em{\sc i\kern-.025em b}\kern-.08em
    T\kern-.1667em\lower.7ex\hbox{E}\kern-.125emX}}
\begin{document}

\title{Quaternion-Based Graph Convolution Network for Recommendation}

\author{
        \IEEEauthorblockN{
            Yaxing Fang\IEEEauthorrefmark{2},
			Pengpeng Zhao\IEEEauthorrefmark{2}, 
			Guanfeng Liu\IEEEauthorrefmark{3}, 
			Yanchi Liu\IEEEauthorrefmark{4}, 
			Victor S. Sheng\IEEEauthorrefmark{6}, 
			Lei Zhao\IEEEauthorrefmark{2}, 
			Xiaofang Zhou\IEEEauthorrefmark{8}
			}
			\IEEEauthorblockA{\IEEEauthorrefmark{2}School of Computer Science and Technology, Soochow University, Suzhou, China}
			\IEEEauthorblockA{\IEEEauthorrefmark{3}Macquarie University, Sydney, Australia, 
			                  \IEEEauthorrefmark{4}Rutgers University, New Jersey, USA}
			\IEEEauthorblockA{\IEEEauthorrefmark{6}Department of Computer Science, Texas Tech University, Lubbock, USA}                 
	    	\IEEEauthorblockA{\IEEEauthorrefmark{8}The Hong Kong University of Science and Technology, Hong Kong SAR, China}
	    	
	    	\IEEEauthorblockA{\IEEEauthorrefmark{2}yxfang2020@stu.suda.edu.cn, \{ppzhao,zhaol\}@suda.edu.cn}
	    	{\IEEEauthorrefmark{3}guanfeng.liu@mq.edu.au}
	    	{\IEEEauthorrefmark{4}yanchi.liu@rutgers.edu}   {\IEEEauthorrefmark{6}victor.sheng@ttu.edu}      {\IEEEauthorrefmark{8}zxf@cse.ust.hk}
		}

\maketitle

\begin{abstract}
Graph Convolution Network (GCN) has been widely applied in recommender systems for its representation learning capability on user and item embeddings. However, GCN is vulnerable to noisy and incomplete graphs, which are common in real world, due to its recursive message propagation mechanism. In the literature, some work propose to remove the feature transformation during message propagation, but making it unable to effectively capture the graph structural features. Moreover, they model users and items in the Euclidean space, which has been demonstrated to have high distortion when modeling complex graphs, further degrading the capability to capture the graph structural features and leading to sub-optimal performance. To this end, in this paper, we propose a simple yet effective Quaternion-based Graph Convolution Network (QGCN) recommendation model. In the proposed model, we utilize the hyper-complex Quaternion space to learn user and item representations and feature transformation to improve both performance and robustness. Specifically, we first embed all users and items into the Quaternion space. Then, we introduce the quaternion embedding propagation layers with quaternion feature transformation to perform message propagation. Finally, we combine the embeddings generated at each layer with the mean pooling strategy to obtain the final embeddings for recommendation. Extensive experiments on three public benchmark datasets demonstrate that our proposed QGCN model outperforms baseline methods by a large margin. 
\end{abstract}

\begin{IEEEkeywords}
Recommender Systems, Collaborative Filtering, Graph Neural Network, Quaternion Embedding
\end{IEEEkeywords}

\section{INTRODUCTION}
Recommender systems have been widely used for alleviating information overload in real-world applications, such as social media \cite{RS_in_social_media}, news \cite{RS_in_news}, videos \cite{RS_in_videos}, and E-commerce \cite{RS_in_E-commerce}. It aims to estimate whether a user will show a preference for an item, based on the user's historical interactions. Among existing recommendation methods, Collaborative Filtering (CF) based models \cite{CF_based_model_1,CF_based_model_2(NeuMF),CF_based_model_3,CF_based_model_4,CF_based_model_5} have shown great performance in user and item representation learning. For example, Matrix factorization \cite{MF_for_Rec} and Neural collaborative filtering model \cite{CF_based_model_2(NeuMF)} are widely used CF models, which embed users and items into the latent space and model the user-item interactions with inner product. 

Recently, GCN-based recommendation models have surged to learn better user and item representations in the user-item bipartite graph. The typical flow can be summarized as follows: 1) Initialize user and item representations by embedding them into the latent space; 2) Use an aggregation function over neighbors of each node to update its representation iteratively; 3) Readout the final representation of each node by combining or concatenating. The paradigm of GCN iterative aggregating feature information from local graph neighbors has been proved to be an efficient way to distill additional information from graph structure and thus improve user and item representation learning. For example, PinSage \cite{PinSage} combines random walk and graph convolutions to learn the embeddings of nodes. GC-MC explores the first-order connectivity between users and items by utilizing only one convolution layer over the user-item bipartite graph. NGCF \cite{NGCF} leverages the message-passing mechanism to obtain high-order connectivity and collaborative signal in the user-item integration graph. LightGCN \cite{LightGCN} removes two components, feature transformation and non-linear activation in NGCF \cite{NGCF}, leading to improvement in training efficiency and generation ability. 

\begin{figure}[t]
\centering
\includegraphics[width=0.9\linewidth]{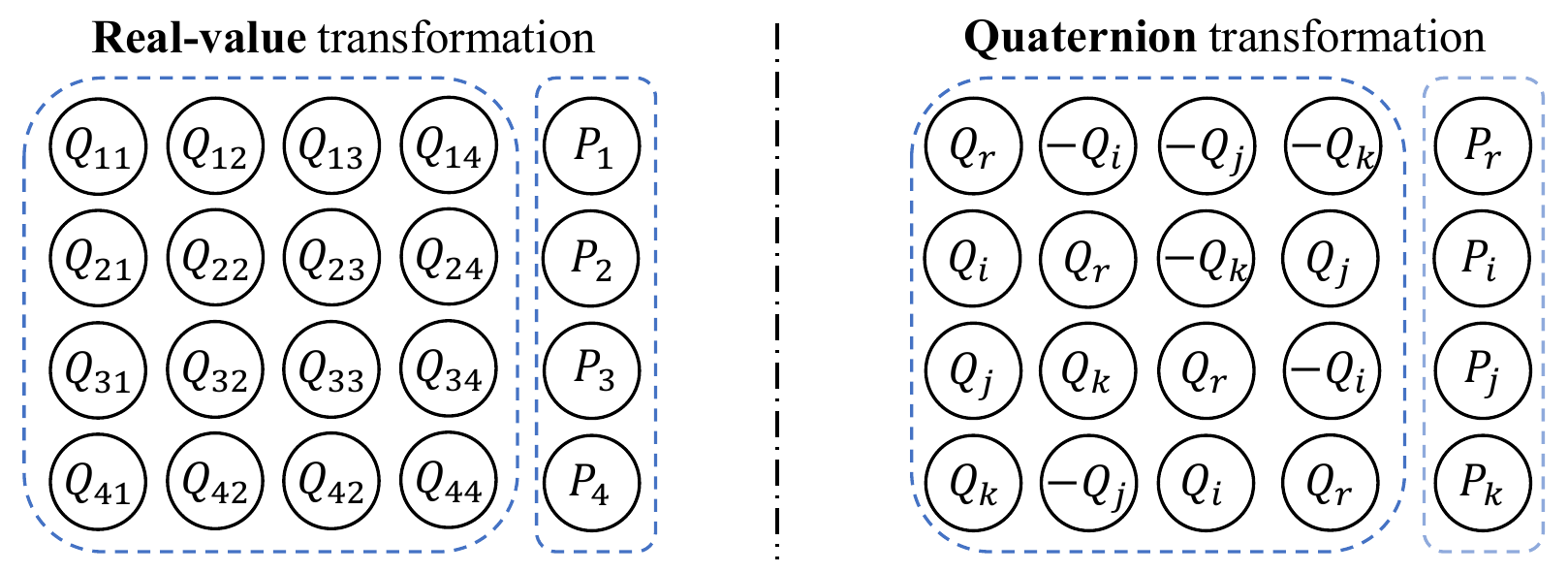}
\caption{Comparison between real-value transformation and quaternion transformation.} 
\label{quaternion transformationl}
% \lyc{pls make the case of captions of tables and figures consistent}
\end{figure}

Despite effectiveness, GCN is still vulnerable to noisy and incomplete graphs, which are common in real-world scenarios, due to its recursive message propagation mechanism \cite{SGCN,vulnerable_to_the_quality_of_graph_1,vulnerable_to_the_quality_of_graph_2}. 
However, some latest GCN-based recommendation models ($e.g.$ LightGCN \cite{LightGCN}) propose to remove the feature transformation during message propagation, but making it unable to effectively capture the graph structural features and become more sensitive to noisy or missing information. Moreover, they model users and items in the Euclidean space, which has been demonstrated to have high distortion when modeling complex graphs \cite{HGCN,HGNN}, further degrading the capability to capture the graph structural features and leading to sub-optimal performance. 

Can we move beyond the Euclidean space to learn better user and item representations and feature transformation, capture the graph structural features more effectively, and thus improve both recommendation performance and model robustness? 
Quaternion space - a hyper-complex vector space, where each quaternion is a hyper-complex number consisting of one real and three imaginary components, has shown great performance in representation learning \cite{QGNN,QuatRE,QKGE}. Hamilton product, which is the multiplication of quaternions, enhances the inter-latent interactions between real and imaginary components of two quaternions, and any slight change in the input quaternion results in an entirely different output, leading to highly expressive computations, and thus the intricate relations are captured more powerfully \cite{Quaternion_expressive}. As shown in Fig.~\ref{quaternion transformationl}, real-value transformation consists of 16 different components $\{ Q_{ij} | i \in [1,4], j \in [1,4]\}$, while Quaternion transformation consists of 4 weighting components $\{ Q_r, Q_i, Q_j, Q_k\}$ due to the wight sharing nature of Hamilton product (\emph{c.f.} Equation~\ref{hamilton product}), leading to up to four times reduction of parameters. 
There has been significant success of quaternion-based methods in various fields. 
For example, \cite{Q_signal_processing} applies a quaternionic Fourier transform and a quaternionic Gabor filter and exploits the symmetries inherent in the quaternion to find differences between subtly varying images. 
\cite{Q_image_classification_1} explores the benefits of generalizing one step further into the Quaternion space and provides the architecture components needed to build deep quaternion networks. 
\cite{Q_image_classification_2} re-designs the basic modules like convolution layers and fully-connected layers in the quaternion domain, which can be used to establish fully-quaternion convolutional neural networks. 
\cite{Q_speech_recognition_2} applies the Quaternion space into recurrent neural network (RNN) and long-short term memory neural network (LSTM) and achieves better performance than the basic model in a realistic application of automatic speech recognition. 
\cite{Q_speech_recognition_3} integrates multiple feature views in quaternion-valued convolutional neural network (QCNN) to be used for sequence-to-sequence mapping with the CTC model. 
\cite{Q_speech_recognition_1} investigates modern quaternion-valued models such as convolutional and recurrent quaternion neural networks in the context of speech recognition.

Recently, there has been some work introducing the Quaternion space into graph representation learning to obtain more expressive graph-level representations \cite{QGNN,QuatRE,QKGE}. For example, \cite{QGNN} generalizes graph neural networks within the Quaternion space for graph classification, node classification, and text classification. \cite{QuatRE,QKGE} introduce more expressive quaternion representations to model entities and relations for knowledge graph embeddings for knowledge graph completion. 
However, there is almost no exploration of the Quaternion space in GCN-based recommendation scenarios. Some challenges during this process remain to be explored. The most crucial one is that: The model should not be designed to be very complex or redundant to better validate the effectiveness of the Quaternion space and for more intuitive comparison. In other words, how to introduce the Quaternion space while keeping the model as simple as possible remains to be considered. 

To this end, in this paper, we propose a simple yet effective \textbf{Q}uaternion-based \textbf{G}raph \textbf{C}onvolution \textbf{N}etwork (QGCN) recommendation model, which improves both performance and robustness. 
Specifically, we first embed all users and items into the Quaternion space with quaternion embeddings. Then, we introduce the quaternion embedding propagation layers with quaternion feature transformation to perform message propagation for aggregating more useful information. Finally, we combine the embeddings generated at each layer with the mean pooling strategy to obtain the final embeddings for recommendation. The quaternion feature transformation enhances the inter-latent interactions between real and imaginary components, enabling it to capture the graph structural features more effectively, distinguish the contribution of different nodes during message propagation, and thus improve both performance and robustness. 
Extensive experiments are conducted on three public benchmark datasets to validate the effectiveness of our proposed QGCN model. Results show that QGCN outperforms the state-of-the-art methods by a large margin, which indicates that it can better learn user and item representations. Besides, with further robustness analysis, we find that the performance of our QGCN model remains steady in various noisy or incomplete graphs, while that of compared state-of-the-art methods declines dramatically. This indicates that our model is more robust and can effectively capture the graph structural features.

We summarize the contributions of this work as follows:
\begin{itemize}
    \item [$\bullet $] To the best of our knowledge, we are the first to introduce the Quaternion space into GCN-based recommendation models.
    \item [$\bullet $] A QGCN model is proposed to model users and items in the Quaternion space and propagate them with quaternion feature transformation, which significantly enhances both recommendation performance and model robustness.
    \item [$\bullet $] We conduct extensive experiments on three public benchmark datasets to evaluate the effectiveness of our proposed model. Experimental results demonstrate that our QGCN model outperforms baseline methods by a large margin, confirming the effectiveness of the quaternion embeddings and quaternion feature transformation. Results of robustness analysis show that our QGCN model is more robust to noisy and incomplete graphs, verifying the effectiveness of the quaternion feature transformation capturing the graph structural features. 
\end{itemize}

\section{PROBLEM STATEMENT}
In this section, we first introduce the notations used in this paper and give a formal problem definition of graph-based collaborative filtering for recommendation. Table~\ref{table of notations} summarizes the notations and the corresponding description. 

We denote the set of users and items as $\mathcal{U}$ and $\mathcal{I}$, and the number of users and items are respectively $M$ and $N$. We construct the user-item interaction matrix $\mathbf{R} \in \mathbb{R}^{M \times N}$ where $\mathbf{R}_{ui}=1$ represents user $u$ has interacted with item $i$. $\mathcal{N}_u$ and $\mathcal{N}_i$ respectively denote the user $u$'s interacted items and the item $i$'s interacted users, respectively. The adjacency matrix $\mathbf{A} \in \mathbb{R}^{(M+N) \times (M+N)}$ is constructed based on the user-item interaction matrix. Then, we define the graph-based collaborative filtering for recommendation as follows. Given the user-item interaction matrix $\mathbf{R}$, our goal is to estimate whether a user $u \in \mathcal{U}$ will show a preference for an item $i \in \mathcal{I}$ based on the user and item embedding generated after $L$ layers' graph convolution. 

\section{PRELIMINARIES}
In this section, we first recap the start-of-the-art framework of GCN-based recommendation models and then cover some necessary background on quaternion before delving into the architecture of our proposed model. 

\subsection{GCN-based Recommendation Models}
Let $\mathbf{e}_u^0$ denote the ID embedding of user $u$ and $\mathbf{e}_i^0$ denote the ID embedding of item $i$.
\subsubsection{NGCF}
NGCF \cite{NGCF} leverages the message-passing mechanism to obtain high-order connectivity and collaborative signal in the user-item integration graph. The message passing strategy and node aggregation is defined as follows:
\begin{equation}
    \begin{split}
        \mathbf{m}^{(l)}_{u} &= \sum \limits _{i \in \mathcal{N}_u} 
        p_{ui}
        \left( 
        \mathbf{W}_1^{(l)} \mathbf{e}_i^{(l-1)} + \mathbf{W}_2^{(l)} (\mathbf{e}_i^{(l-1)} \odot \mathbf{e}_u^{(l-1)})
        \right), \\
        \mathbf{m}^{(l)}_{u} &= \sum \limits _{i \in \mathcal{N}_u} 
        p_{ui}
        \left( 
        \mathbf{W}_1^{(l)} \mathbf{e}_i^{(l-1)} + \mathbf{W}_2^{(l)} (\mathbf{e}_u^{(l-1)} \odot \mathbf{e}_i^{(l-1)})
        \right),
    \end{split}
\end{equation}
where $\mathbf{m}^{(l)}_{u}$ and $\mathbf{m}^{(l)}_{u}$ respectively denote the message propagated from user $u$'s and item $i$'s neighbors; $p_{ui}$ is set to the graph Laplacian norm $1 / \sqrt{|\mathcal{N}_u||\mathcal{N}_i|}$, where $|\mathcal{N}_u|$ and $|\mathcal{N}_i|$ respectively denote user $u$'s interacted items and item $i$'s interacted users; $\mathbf{W}_1^{(l)}$ and $\mathbf{W}_2^{(l)}$ are the trainable transformation matrices.

Then the user and item embedding are updated by the sum of the node embedding itself and its neighbors with an activation function LeakyReLU:
\begin{equation}
    \begin{split}
        \mathbf{e}^{(l)}_{u} &= LeakyReLU(\mathbf{W}_1^{(l)} \mathbf{e}_u^{(l-1)} + \mathbf{m}^{(l)}_{u}), \\
        \mathbf{e}^{(l)}_{i} &= LeakyReLU(\mathbf{W}_1^{(l)} \mathbf{e}_i^{(l-1)} + \mathbf{m}^{(l)}_{i}).
    \end{split}
\end{equation}

NGCF adopts the concatenation strategy that the representations generated at each layer are concatenated as the final node representation:
\begin{equation}
    \mathbf{e}_{u} = \mathbf{e}^{(0)}_{u} \| \dots \| \mathbf{e}^{(L)}_{u}, \quad
    \mathbf{e}_{i} = \mathbf{e}^{(0)}_{i} \| \dots \| \mathbf{e}^{(L)}_{i}.
\end{equation}

\subsubsection{LightGCN}
LightGCN \cite{LightGCN} removes two components, feature transformation and non-linear activation in NGCF. It not only simplifies the model itself but also leads to improvement in training efficiency and generation ability. The embedding propagation is defined as follows:
\begin{equation}
    \begin{split}
        \mathbf{e}^{(l)}_{u} &= \sum \limits _{i \in \mathcal{N}_u} \frac{1}{\sqrt{|\mathcal{N}_u||\mathcal{N}_i|}} \mathbf{e}^{(l-1)}_{i}, \\
        \mathbf{e}^{(l)}_{i} &= \sum \limits _{u \in \mathcal{N}_i}     \frac{1}{\sqrt{|\mathcal{N}_i||\mathcal{N}_u|}} \mathbf{e}^{(l-1)}_{u}.
    \end{split}
\end{equation}

% table of notations
\begin{table}[t]
\renewcommand\arraystretch{1.4}
\centering
    \caption{Table of notations.}
    \label{table of notations}
    % \resizebox{0.8\linewidth}{!}{}
    \resizebox{0.99\linewidth}{!}{
    \begin{tabular}{c|l}
        \hline
        \textbf{Notations}  &  \textbf{Descriptions} \\
        \hline \hline
        $\mathcal{U}$, $\mathcal{I}$  & the set of users and items \\
        \hline
        $M$, $N$ & the number of users and items \\ 
        \hline
        $\mathbf{R} \in \mathbb{R}^{M \times N}$ & the user-item interaction matrix \\
        \hline
        \multirow{2}{*}{$\mathbf{R}_{ui}$} & $\mathbf{R}_{ui}=1$ represents user $u$ \\
        & has interacted with item $i$ \\
        \hline
        \multirow{2}{*}{$\mathbf{A} \in \mathbb{R}^{(M+N) \times (M+N)}$} & the adjacency matrix of \\
        & the user-item interaction matrix \\
        \hline
        $|V|$, $|E|$ & the number of nodes and edges \\
        \hline
        $\mathcal{N}_u$ & the user $u$'s interacted items \\
        \hline
        $\mathcal{N}_i$ & the item $i$'s interacted users \\
        \hline
        $d$ & the quaternion dimension \\
        \hline
        $e_u^{l, Q} \in \mathbb{H}^{d}$ & the user embedding at layer $l$ \\
        \hline
        $e_u^{l, Q} \in \mathbb{H}^{d}$ & the item embedding at layer $l$ \\
        \hline
        \multirow{2}{*}{$\mathbf{W}^{l,Q} \in \mathbb{H}^{d \times d}$} & the quaternion feature \\
        &transformation matrix at layer $l$ \\
        \hline
        $L$ & the number of graph convolution layers \\
        \hline
        $e_u^{Q} \in \mathbb{H}^{d}$ & the final representation of user $u$ \\
        \hline
        $e_i^{Q} \in \mathbb{H}^{d}$ & the final representation of item $i$ \\
        \hline
        $\|$ & vector concatenation \\
        \hline
        $\odot$ & element-wise product \\
        \hline
        $\otimes$ & Hamilton product \\
        \hline
    \end{tabular}
    }
\end{table}

Different from NGCF, LightGCN adopts weighted sum strategy to aggregate the representations at each layer:
\begin{equation}
    \begin{split}
        \mathbf{e}_{u} = \sum \limits _{k=0} ^L \lambda _k \mathbf{e}^{(k)}_{u},
        \quad
        \mathbf{e}_{i} = \sum \limits _{k=0} ^L \lambda _k \mathbf{e}^{(k)}_{i},
    \end{split}
\end{equation}
where $\lambda _k \geq 0$ denotes the importance of the $k$-th layer embedding for the final node embedding.

After obtaining the final representations of nodes, the inner product is conducted to estimate the user $u$'s preference towards the target item $i$:
\begin{equation}
    \hat{y_{ui}} = {\mathbf{e}_{u}}^ \mathrm{T} \mathbf{e}_{i}.
\end{equation}

The $\emph{Bayesian Personalized Ranking}$ (BPR) loss \cite{BPR-loss} is employed in both NGCF and LightGCN to optimize the model parameters, \emph{i.e.} minimizing the following loss function:
\begin{equation}
    Loss = \sum \limits _{u=1} ^M \sum _{i \in \mathcal{N}_u} \sum _{j \notin \mathcal{N}_u} - \ln \sigma(\hat{y}_{ui} - \hat{y}_{uj}) + \lambda \Vert \mathbf{\Theta} \Vert _2 ^2,
\end{equation}
where $\mathcal{N}_u$ denotes user $u$'s interacted items; $\sigma$ is the sigmoid function; $\lambda$ represents the regularization weight and $\mathbf{\Theta}$ denotes model parameters.

\subsection{Quaternion}
\subsubsection{Quaternion} 
A quaternion $\emph{Q} \in \mathbb{H}$ is a hyper-complex number consisting of one real part and three imaginary parts defined as:
\begin{equation}
    Q = Q_r + Q_i\mathbf{i} + Q_j \mathbf{j} + Q_k \mathbf{k},
\end{equation}
where $Q_r, Q_i, Q_j, Q_k \in \mathbb{R}$, and $\mathbf{i}$, $\mathbf{j}$, $\mathbf{k}$ are imaginary units, satisfying the following rule: 
\begin{equation}
    \mathbf{i}^2 = \mathbf{j}^2 = \mathbf{k}^2 = \mathbf{i}\mathbf{j}\mathbf{k} = -1. 
\end{equation}
% $\mathbf{i}^2 = \mathbf{j}^2 = \mathbf{k}^2 = \mathbf{i}\mathbf{j}\mathbf{k} = -1$. 

Corresponding to the definition of quaternion, the $n$-dimensional vector form of quaternion $\boldsymbol {Q} \in \mathbb{H}^n$ is defined as:
\begin{equation}
    \boldsymbol {Q} = \boldsymbol {Q_r} + \boldsymbol {Q_i} \mathbf{i} +
    \boldsymbol {Q_j} \mathbf{j} + \boldsymbol {Q_k} \mathbf{k},
\end{equation}
where $\boldsymbol {Q_r}, \boldsymbol {Q_i}, \boldsymbol {Q_j}, \boldsymbol {Q_k} \in \mathbb{R}^n$.

\subsubsection{Quaternion Addition}
The addition of two quaternions $\emph{Q}$ and $\emph{P}$ is defined as:
\begin{equation}
    Q + P = (Q_r+P_r) + (Q_i+P_i)\mathbf{i} + (Q_j+P_j)\mathbf{j} + (Q_k+P_k)\mathbf{k},
\end{equation}

\subsubsection{Quaternion Inner Product}
The inner product of two quaternions $\emph{Q}$ and $\emph{P}$ is defined as:
\begin{equation}
    Q \cdot P = Q_r \cdot P_r + Q_i \cdot P_i + Q_j \cdot P_j + Q_k \cdot P_k.
\end{equation}

\subsubsection{Hamilton Product}
The quaternion product of two quaternions $\emph{Q}$ and $\emph{P}$ is defined as:

\begin{equation}
\label{normal hamilton product}
\centering
    \begin{split}
        Q \otimes P &= (Q_r P_r-Q_i P_i-Q_j P_j-Q_k P_k) \\
        &+ (Q_i P_r+Q_r P_i-Q_k P_j+Q_j P_k) \mathbf{i}  \\
        &+ (Q_j P_r+Q_k P_i+Q_r P_j-Q_i P_k) \mathbf{j}  \\
        &+ (Q_k P_r-Q_j P_i+Q_i P_j+Q_r P_k) \mathbf{k}.
    \end{split}
\end{equation}

We further simplify the result of Hamilton product above into matrix form as follows:

\begin{equation}
\centering
\label{hamilton product}
    \begin{bmatrix}
    1 \\ \mathbf{i} \\ \mathbf{j} \\ \mathbf{k}
    \end{bmatrix} ^ \mathrm{T}
    \begin{bmatrix}
    Q_{r} & -Q_{i} & -Q_{j} & -Q_{k} \\
    Q_{i} & Q_{r} & -Q_{k} & Q_{j} \\
    Q_{j} & Q_{k} & Q_{r} & -Q_{i} \\
    Q_{k} & -Q_{j} & Q_{i} & Q_{r}
    \end{bmatrix}
    \begin{bmatrix}
    P_{r} \\ P_{i} \\ P_{j} \\ P_{k}
    \end{bmatrix}.
\end{equation}

\section{METHODOLOGY}

\begin{figure*}[t]
\centering
\includegraphics[width=0.8\textwidth]{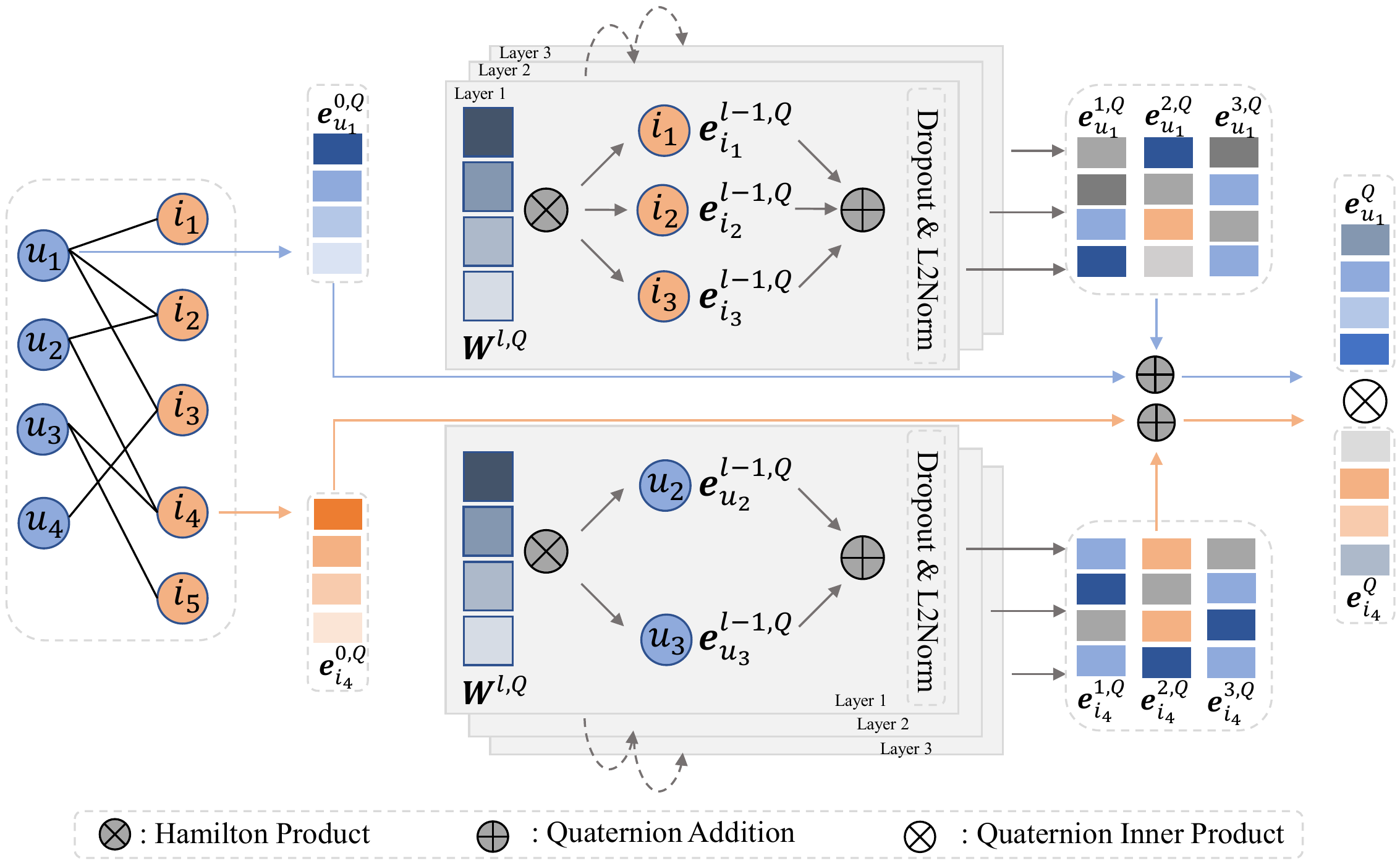}
\caption{The architecture of our proposed QGCN model which is formed by Quaternion Embedding Layer, Quaternion Embedding Propagation Layers and Prediction Layer.}
\label{QGCN_model}
\end{figure*}

In this section, we present our proposed QGCN model. As illustrated in Fig.~\ref{QGCN_model}, the model contains three main components: Quaternion Embedding Layer, Quaternion Embedding Propagation Layers, and Prediction Layer.

\subsection{Quaternion Embedding Layer}
Firstly, we embed all the users and items into the Quaternion space. For each user $u \in \mathcal{U}$, we represent it with a quaternion ID embedding $\mathbf{e}^{0, Q}_{u} \in \mathbb{H}^d$, where $d$ represents the quaternion dimension. And the same for item quaternion ID embeddings, each item $i \in \mathcal{I}$ is initialized with a quaternion ID embedding $\mathbf{e}^{0, Q}_{i} \in \mathbb{H}^d$. The initial quaternion ID embedding for users and items can be defined as follows:
\begin{equation}
\begin{split}
\mathbf{e}^{0, Q}_{u_1} &= \mathbf{e}^0_{u_1, r} + \mathbf{e}^0_{u_1, i}\mathbf{i} + \mathbf{e}^0_{u_1, j}\mathbf{j} + \mathbf{e}^0_{u_1, k}\mathbf{k}, \\
&......\\
\mathbf{e}^{0, Q}_{u_M} &= \mathbf{e}^0_{u_M, r} + \mathbf{e}^0_{u_M, i}\mathbf{i} + \mathbf{e}^0_{u_M, j}\mathbf{j} + \mathbf{e}^0_{u_M, k}\mathbf{k}, \\
\mathbf{e}^{0, Q}_{i_1} &= \mathbf{e}^0_{i_1, r} + \mathbf{e}^0_{i_1, i}\mathbf{i} + \mathbf{e}^0_{i_1, j}\mathbf{j} + \mathbf{e}^0_{i_1, k}\mathbf{k}, \\
&......\\
\mathbf{e}^{0, Q}_{i_N} &= \mathbf{e}^0_{i_N, r} + \mathbf{e}^0_{i_N, i}\mathbf{i} + \mathbf{e}^0_{i_N, j}\mathbf{j} + \mathbf{e}^0_{i_N, k}\mathbf{k}, \\
\end {split}
\end{equation}
% \begin{equation}
%     \begin{split}
%     \mathbf{e}^{0, Q}_{i_1} &= \mathbf{e}^0_{i_1, r} + \mathbf{e}^0_{i_1, i}\mathbf{i} + \mathbf{e}^0_{i_1, j}\mathbf{j} + \mathbf{e}^0_{i_1, k}\mathbf{k}, \\
%     &......\\
%     \mathbf{e}^{0, Q}_{i_N} &= \mathbf{e}^0_{i_N, r} + \mathbf{e}^0_{i_N, i}\mathbf{i} + \mathbf{e}^0_{i_N, j}\mathbf{j} + \mathbf{e}^0_{i_N, k}\mathbf{k}, \\
%     \end {split}
% \end{equation}
where $\mathbf{e}^0_{u_m, r}, \mathbf{e}^0_{u_m, i}, \mathbf{e}^0_{u_m, j}, \mathbf{e}^0_{u_m, k} \in \mathbb{R}^d$, $\forall m \in \{1, 2, \dots, M\}$ and $\mathbf{e}^0_{i_n, r}, \mathbf{e}^0_{i_n, i}, \mathbf{e}^0_{i_n, j}, \mathbf{e}^0_{i_n, k} \in \mathbb{R}^d$, $\forall n \in \{1, 2, \dots, N\}$. $M$, $N$ respectively denote the number of users and items.

\subsection{Quaternion Embedding Propagation Layers}
\subsubsection{Quaternion Embedding Propagation}
Next, we perform message propagation within the Quaternion Embedding Propagation Layers with quaternion feature transformation. As mentioned above, we argue that removing the feature transformation during message propagation makes it unable to effectively capture the graph structural features and become more sensitive to noisy or missing information, further degrading the model performance. So in this part, we introduce the feature transformation in the Quaternion space at each layer for message propagation to aggregate more useful information. In order to prove our quaternion feature transformation to be valid more intuitively, we adopt the simple message propagation procedure like the vanilla GCN \cite{vanilla-GCN} without the non-linear activation function, only involving the user and item embeddings and the quaternion transformation matrices. We generate the quaternion transformation matrix at layer $l$ as follows:
\begin{equation}
\mathbf{W}^{l, Q} = \mathbf{W}^l_{r} + \mathbf{W}^l_{i}\mathbf{i} + \mathbf{W}^l_{j}\mathbf{j} + \mathbf{W}^l_{k}\mathbf{k},
\end{equation}
where $\mathbf{W}^l_{r}, \mathbf{W}^l_{i}, \mathbf{W}^l_{j}, \mathbf{W}^l_{k} \in \mathbb{R}^{d \times d}$.

Thus, our quaternion embedding propagation rule in QGCN is defined as:
\begin{equation}
\label{quaternion embedding propagation}
    \begin{split}
        \mathbf{e}^{l,Q}_{u} &= \sum \limits _{i \in \mathcal{N}_u} \frac{1}{\sqrt{|\mathcal{N}_u||\mathcal{N}_i|}} \mathbf{W}^{l,Q} \otimes \mathbf{e}^{l-1,Q}_{i}, \\
        \mathbf{e}^{l,Q}_{i} &= \sum \limits _{u \in \mathcal{N}_i} \frac{1}{\sqrt{|\mathcal{N}_i||\mathcal{N}_u|}} \mathbf{W}^{l,Q} \otimes \mathbf{e}^{l-1,Q}_{u},
    \end{split}
\end{equation}
where $\mathbf{e}^{l,Q}_{u}$ and $\mathbf{e}^{l,Q}_{i}$ respectively represent user $u$'s quaternion embedding and item $i$'s quaternion embedding after $l$ layers propagation; $1 / \sqrt{|\mathcal{N}_u||\mathcal{N}_i|}$ is the symmetric normalization term following the vanilla GCN \cite{vanilla-GCN}, designed to avoid the scale of embeddings increasing with graph convolution operations, where $\mathcal{N}_u$ and $\mathcal{N}_i$ respectively denote the user $u$'s interacted items and the item $i$'s interacted users; $\mathbf{W}^{l,Q} \in \mathbb{H}^{d \times d}$ is the quaternion feature transformation matrix at layer $l$; $\otimes$ denotes Hamilton product.

To facilitate the implementation of the quaternion embedding propagation, we derive the Hamilton product $\otimes$ between $\mathbf{W}^{l, Q}$ and $\mathbf{e}^{l-1,Q}_{u}$ in Equation~\ref{quaternion embedding propagation} as follows (\emph{c.f.} Equation~\ref{hamilton product}):
\begin{equation}
    % \mathbf{W}^{l,Q} \otimes \mathbf{e}^{l-1,Q}_{u} = 
    \begin{bmatrix}
    1 \\ \mathbf{i} \\ \mathbf{j} \\ \mathbf{k}
    \end{bmatrix} ^ \mathrm{T}
    \begin{bmatrix}
    \mathbf{W}^l_{r} & -\mathbf{W}^l_{i} & -\mathbf{W}^l_{j} & -\mathbf{W}^l_{k} \\
    \mathbf{W}^l_{i} & \mathbf{W}^l_{r} & -\mathbf{W}^l_{k} & \mathbf{W}^l_{j} \\
    \mathbf{W}^l_{j} & \mathbf{W}^l_{k} & \mathbf{W}^l_{r} & -\mathbf{W}^l_{i} \\
    \mathbf{W}^l_{k} & -\mathbf{W}^l_{j} & \mathbf{W}^l_{i} & \mathbf{W}^l_{r}
    \end{bmatrix}
    \begin{bmatrix}
    \mathbf{e}^{l-1}_{u, r} \\ \mathbf{e}^{l-1}_{u, i} \\ \mathbf{e}^{l-1}_{u, j} \\ \mathbf{e}^{l-1}_{u, k}
    \end{bmatrix}.
\label{user update in normal form}
\end{equation}

Similarly, the result of Hamilton product $\otimes$ between $\mathbf{W}^{l, Q}$ and $\mathbf{e}^{l-1,Q}_{i}$ can be derived as follows:
\begin{equation}
    \begin{bmatrix}
    1 \\ \mathbf{i} \\ \mathbf{j} \\ \mathbf{k}
    \end{bmatrix} ^ \mathrm{T}
    \begin{bmatrix}
    \mathbf{W}^l_{r} & -\mathbf{W}^l_{i} & -\mathbf{W}^l_{j} & -\mathbf{W}^l_{k} \\
    \mathbf{W}^l_{i} & \mathbf{W}^l_{r} & -\mathbf{W}^l_{k} & \mathbf{W}^l_{j} \\
    \mathbf{W}^l_{j} & \mathbf{W}^l_{k} & \mathbf{W}^l_{r} & -\mathbf{W}^l_{i} \\
    \mathbf{W}^l_{k} & -\mathbf{W}^l_{j} & \mathbf{W}^l_{i} & \mathbf{W}^l_{r}
    \end{bmatrix}
    \begin{bmatrix}
    \mathbf{e}^{l-1}_{i, r} \\ \mathbf{e}^{l-1}_{i, i} \\ \mathbf{e}^{l-1}_{i, j} \\ \mathbf{e}^{l-1}_{i, k}
    \end{bmatrix}.
\label{item update in normal form}
\end{equation}

\subsubsection{Dropout and L2Norm}
Dropout drops the units of the neural networks with a certain probability during the training process, which proves to be an effective way to prevent neural networks from overfitting \cite{dropout_origin_1, dropout_origin_2}. Motivated by the previous work of introducing dropout into graph convolutional network \cite{dropout_in_GCN} and GCN-based recommendation models \cite{NGCF}, we apply dropout to the user and item embeddings at each layer $l$ with a certain dropout rate $p$, which is one of the critical hyper-parameters to be tuned. Then, we perform L2 Normalization function on them for training speed and stability. We summarize the dropout and L2 normalization as follows:
\begin{equation}
\label{dropour and l2 norm}
    \begin{split}
        \mathbf{e}^{l,Q}_{u} &= L2Norm \left(Dropout(\mathbf{e}^{l,Q}_{u}) \right), \\
        \mathbf{e}^{l,Q}_{i} &= L2Norm \left(Dropout(\mathbf{e}^{l,Q}_{i}) \right).
    \end{split}
\end{equation}

\subsubsection{Quaternion Propagation Rule in Matrix Form.}
To better facilitate the implementation of our QGCN model, we provide the quaternion embedding propagation rule in matrix form. 
As defined in Table~\ref{table of notations}, we denote the the user-item interaction matrix as $\mathbf{R} \in \mathbb{R}^{M \times N}$, where $M$ and $N$ denote the number of users and items respectively and each element in $\mathbf{R}$ denotes the interaction, that if user $u$ has interacted with item $i$, then $\mathbf{R}_{ui}$ is set to 1, otherwise 0. Then, the adjacency matrix of the user-item graph $\mathbf{A} \in \mathbb{R}^{(M+N) \times (M+N)}$can be generated as:
\begin{equation}
    \mathbf{A} = 
    \begin{pmatrix}
    \mathbf{0} & \mathbf{R}   \\
    \mathbf{R} ^ \mathrm{T} & \mathbf{0}
    \end{pmatrix}.
\end{equation}
Next, we can obtain the diagonal matrix $\mathbf{D} \in \mathbb{R}^{(M+N) \times (M+N)}$ correspondingly, where each diagonal element $D_{ii}$ denotes the number of nonzero nodes in the $i$-th row vector of the adjacency matrix $\mathbf{A}$.

Then, we generate the Laplacian matrix $\mathcal{L} = D^{-\frac{1}{2}} A D^{-\frac{1}{2}}$. As mentioned above, we derive the Hamilton product to facilitate the implementation of the quaternion embedding propagation in Equation~\ref{user update in normal form} and Equation~\ref{item update in normal form}. Thus, we obtain the quaternion propagation in matrix form as:
\begin{equation}
    \mathbf{E}^{l} = \mathcal{L} \mathbf{E}^{l-1,Q} \mathbf{W}^{l},
\end{equation}
\begin{equation}
    \mathbf{W}^{l} = 
    \begin{bmatrix}
    \mathbf{W}^l_{r} & -\mathbf{W}^l_{i} & -\mathbf{W}^l_{j} & -\mathbf{W}^l_{k} \\
    \mathbf{W}^l_{i} & \mathbf{W}^l_{r} & -\mathbf{W}^l_{k} & \mathbf{W}^l_{j} \\
    \mathbf{W}^l_{j} & \mathbf{W}^l_{k} & \mathbf{W}^l_{r} & -\mathbf{W}^l_{i} \\
    \mathbf{W}^l_{k} & -\mathbf{W}^l_{j} & \mathbf{W}^l_{i} & \mathbf{W}^l_{r}
    \end{bmatrix} ^ \mathrm{T},
\end{equation}
where $\mathbf{E}^{l,Q} \in \mathbb{H}^{(M+N)\times d}$ denote the embedding look-up table at layer $l$, $\mathbf{E}^{l,Q} = (\mathbf{e}_{u_1}^{l,Q}, \dots, \mathbf{e}_{u_M}^{l,Q}, \mathbf{e}_{i_1}^{l,Q}, \dots, \mathbf{e}_{i_N}^{l,Q})$; $\mathbf{W}^{l,Q} \in \mathbb{H}^{d \times d}$ denotes the quaternion feature transformation matrix at layer $l$.

After the quaternion embedding propagation, we apply dropout and L2 normalization to them:
\begin{equation}
    \mathbf{E}^{l,Q} = L2Norm \left(Dropout(\mathbf{E}^{l,Q}) \right).
\end{equation}

\subsection{Prediction Layer}
After the above $L$ layers' quaternion embedding propagation, dropout and L2 normalization, we obtain $L+1$ representations for each user $u$ and item $i$, including the user embedding initialized at quaternion embedding layer, ${\mathbf{e}^{0,Q}_{u}}$ and user representations generated at each layer during propagation, $\{\mathbf{e}^{1,Q}_{u}, \mathbf{e}^{2,Q}_{u}, \dots, \mathbf{e}^{L,Q}_{u}\}$. And the same for item $i$, we obtain $L+1$ item representations which consist of $\{\mathbf{e}^{0,Q}_{i}, \{\mathbf{e}^{1,Q}_{i}, \mathbf{e}^{2,Q}_{i}, \dots, \mathbf{e}^{L,Q}_{i}\}\}$. Since the output of different layers expresses different connections, utilizing the representations of all layers seems like an effective method for GCN-based models. Readout function is the method to obtain the final node representation, $e.g.$ Max, Sum, Concat, Mean pooling, which are the most primitive and simple pooling methods. Specifically, Max, Sum, Mean pooling respectively take the maximum, sum, mean value of the corresponding position of representations generated at each layer, and Concat concatenates representations generated at each layer. We summarize these readout functions as follows:
% \begin{equation}
%     Max/Sum/Mean = Max/Sum/Mean \{\mathbf{e}^{l}_{u} \} _{l=0} ^L.
% \end{equation}
% \begin{equation}
%     Concat = \mathbf{e}^{0}_{u} \| \mathbf{e}^{1}_{u} \| \dots \| \mathbf{e}^{L}_{u}.
% \end{equation}
% Readout function is the method to obtain the final node representation. 
% Max, Sum, Concat, Mean pooling, which are the most primitive and simple pooling methods and can be summarized as follows:

Max pooling takes the maximum value of the corresponding position of representations at each layer:
\begin{equation}
    Max = Max \{\mathbf{e}^{l}_{u} \} _{l=0} ^L.
\end{equation}

Sum pooling sums over value of the corresponding position of representations at each layer:
\begin{equation}
    Sum = Sum \{\mathbf{e}^{l}_{u} \} _{l=0} ^L.
\end{equation}

Concat concatenates representations at each layer:
\begin{equation}
    Concat = \mathbf{e}^{0}_{u} \| \mathbf{e}^{1}_{u} \| \dots \| \mathbf{e}^{L}_{u}.
\end{equation}

Mean pooling takes the mean value of the corresponding position of representations at each layer:
\begin{equation}
    Mean = Mean \{\mathbf{e}^{l}_{u} \} _{l=0} ^L.
\end{equation}

Since we generate user and item representations in the form of quaternion hyper-complex vector, we first concatenate the real and imaginary components of the node embeddings and then apply the original pooling methods as follows:
% \begin{equation}
%     \begin{split}
%         \mathbf{e}^{l}_{u} &= \text{Concat} \{\mathbf{e}^{l,Q}_{u,r}, \mathbf{e}^{l,Q}_{u,i} \mathbf{e}^{l,Q}_{u,j}, \mathbf{e}^{l,Q}_{u,k}\}, \\
%         \mathbf{e}^{*}_{u} &= \text{Readout} \{\mathbf{e}^{l}_{u} \} _{l=1} ^L.
%     \end{split}
% \end{equation}
\begin{equation}
    \mathbf{e}^{l}_{u} = Concat \{\mathbf{e}^{l,Q}_{u,r}, \mathbf{e}^{l,Q}_{u,i} \mathbf{e}^{l,Q}_{u,j}, \mathbf{e}^{l,Q}_{u,k}\},
\end{equation}
\begin{equation}
    \mathbf{e}^{*}_{u} = Readout \{\mathbf{e}^{l}_{u} \} _{l=1} ^L,
\end{equation}
where \emph{Readout} is the readout function (\emph{i.e.} Max, Sum, Concat, Mean pooling) applied on the node embeddings generated at each layer. We further conduct experiments and investigate the influence of the readout function applied to our model in the ablation study part.

After generating the final user and item embeddings, we predict by the inner product of user $u$ and item $i$:
\begin{equation}
    \hat{y_{ui}} = {\mathbf{e}^{*}_{u}}^ \mathrm{T} \mathbf{e}^{*}_{i}.
\end{equation}

\subsection{Optimization}
We adopt $\emph{Bayesian Personalized Ranking}$ (BPR) loss \cite{BPR-loss}, which encourages the observed interactions to achieve higher scores than the unobserved ones. The objective function for our QGCN model is as follows:
\begin{equation}
\label{loss function}
    Loss = \sum \limits _{u=1} ^M \sum _{i \in \mathcal{N}_u} \sum _{j \notin \mathcal{N}_u} - \ln \sigma(\hat{y}_{ui} - \hat{y}_{uj}) 
    + \lambda \Vert \mathbf{\Theta} \Vert _2 ^2 ,
\end{equation}
where $\mathcal{N}_u$ denotes user $u$'s interacted items; $\sigma$ is the sigmoid function; $\lambda$ represents the regularization weight, which is $L_2$ regularization to prevent overfitting; $\mathbf{\Theta} = \left \{\{\mathbf{e}_u^{0,Q}\}_{u \in \mathcal{U}}, \{\mathbf{e}_i^{0,Q}\}_{i \in \mathcal{I}}, \{\mathbf{W}^{l,Q}\}_{l \in [1,L]}\right\}$ denotes all trainable parameters of QGCN. The mini-batch Adam \cite{Adam} is adopted to optimize the prediction model and update the model parameters. In particular, for a batch of randomly sampled triples $\{(u,i,j) | i \in \mathcal{N}_u,j \notin \mathcal{N}_u \}$, their representations can be obtained by the propagation rules and then the model parameters are updated by using the gradients of the loss function.

\begin{algorithm}[h]
		\caption{Quaternion Graph Convolution Algorithm}
		\hspace*{0.02in} {\bf Input:}
		\hspace*{0.02in} User-item interaction matrix $\mathbf{R}$, 
		the number of graph convolution layers $L$, 
		the initialized user embedding $\mathbf{e}^{0, Q}_{u}$ and item embeddings $\mathbf{e}^{0, Q}_{i}$, 
		the quaternion transformation matrix $\mathbf{W}^{l,Q}$.
		\begin{algorithmic}[1]
		    \For{layer l = 1 $\rightarrow$ L}
               \For{each user $u$, item $i$}
		        \State  $\mathbf{e}^{l,Q}_{u} = \sum \limits _{i \in \mathcal{N}_u} \frac{1}{\sqrt{|\mathcal{N}_u||\mathcal{N}_i|}} \mathbf{W}^{l,Q} \otimes \mathbf{e}^{l-1,Q}_{i}$, \\
                \State  $\mathbf{e}^{l,Q}_{i} = \sum \limits _{u \in \mathcal{N}_i} \frac{1}{\sqrt{|\mathcal{N}_i||\mathcal{N}_u|}} \mathbf{W}^{l,Q} \otimes \mathbf{e}^{l-1,Q}_{u}$;
		        \State  $\mathbf{e}^{l,Q}_{u} = L2Norm \left(Dropout(\mathbf{e}^{l,Q}_{u}) \right)$,
		        \State  $\mathbf{e}^{l,Q}_{i} = L2Norm \left(Dropout(\mathbf{e}^{l,Q}_{i}) \right)$;
	            \EndFor
            \EndFor
			\State $\mathbf{e}^{l}_{u} = Concat \{\mathbf{e}^{l,Q}_{u,r}, \mathbf{e}^{l,Q}_{u,i} \mathbf{e}^{l,Q}_{u,j}, \mathbf{e}^{l,Q}_{u,k}\}$
            \State $\mathbf{e}^{*}_{u} = Mean \{\mathbf{e}^{l}_{u} \} _{l=1} ^L$
		\end{algorithmic}
		\hspace*{0.02in} {\bf Output:}
		Final user embedding $\mathbf{e}^{*}_{u}$ and item embedding $\mathbf{e}^{*}_{i}$ for recommendation.
	\label{alg 1}
\end{algorithm}

\subsection{Complexity Analysis}
As defined in Table~\ref{table of notations}, $\mathbf{R}$, $|V|$ and $|E|$, $L$ and $d$ respectively represent the user-item interaction matrix, the number of nodes and edges, the number of graph convolution layers, and the quaternion dimension. 
\subsubsection{Time Complexity}
The time complexity of our model is mainly in the following three parts, adjacency matrix, graph convolution, and BPR loss. For the adjacency matrix, the time complexity is $\mathcal{O}(|E|)$ that we set each element $\mathbf{R}_{ui}=1$ in user-item interaction matrix $\mathbf{R}$ if user $u$ has interacted with item $i$. For the graph convolution, the quaternion embedding propagation has computation complexity $\mathcal{O}(L|E|d^2)$. For the BPR loss, the time complexity is $\mathcal{O}(|E|d)$. Therefore, the overall time complexity of our model is $\mathcal{O}(|E|+L|E|d^2+|E|d)$.

\subsubsection{Space Complexity}
The space complexity of our model is mainly in the user and item embeddings and the quaternion transformation matrix at each layer. Therefore, the overall space complexity of our model is $\mathcal{O}(|V|d+Ld^2)$. 

\section{EXPERIMENTS}
In this section, we first briefly describe the datasets and our experimental settings, including evaluation metrics, baselines, and parameter settings. Then, we conduct a detailed comparison with LightGCN \cite{LightGCN} and some state-of-the-art baseline methods, followed by the experimental results and our detailed analysis. Moreover, we perform a robustness analysis to explore the robustness of our QGCN model to noisy and incomplete graphs. Besides, ablation studies are performed to investigate the influence of readout function and different components of our QGCN model on the model performance. Finally, we discuss the impact of the critical hyper-parameters on the final results. Specifically, we conduct experiments to try to answer the following research questions:
\begin{itemize}%
    \item [$\bullet $] \textbf{RQ1} How does our proposed QGCN model perform compared with the state-of-the-art baselines?
    \item [$\bullet $] \textbf{RQ2} How can QGCN alleviate the problem of noisy or incomplete graphs? 
    \item [$\bullet $] \textbf{RQ3} What is the influence of readout function, quaternion embedding and quaternion weight matrices on the model performance? 
    \item [$\bullet $] \textbf{RQ4} How do the key hyper-parameters, such as dropout rate and regularization affect the effectiveness of QGCN? 
\end{itemize}

\begin{table}[t]
\renewcommand\arraystretch{1.4}
\centering
    \caption{Statistics of the experimented data.}
    \label{datasets-table}
    \resizebox{0.99\linewidth}{!}{
    \begin{tabular}{c|c|c|c|c}
        \hline
        Dataset & \#Users & \#Items & \#Interactions & \#Density \\
        \hline \hline
        Yelp2018 & 31668 & 38048 & 1561406 & 0.00130 \\
        Amazon-Book & 52643 & 91599 & 2984108 & 0.00062 \\
        Kindle-Store & 68223 & 61934 & 982618 & 0.00023 \\
       \hline
    \end{tabular}
    }
\end{table}

\subsection{Datasets}
To evaluate the effectiveness of QGCN, we conduct experiments on three benchmark datasets: Yelp2018 \cite{LightGCN}, Amazon-Book \cite{LightGCN}, and Amazon-Kindle-Store \cite{IMP-GCN}, which are publicly available. The first dataset is the 2018 edition Yelp\footnote{https://www.yelp.com/dataset} released by the Yelp challenge. The last two datasets are two widely used datasets for product recommendation from Amazon review \footnote{https://jmcauley.ucsd.edu/data/amazon/}. 

Following the general dataset settings in previous recommendation methods, we filter users and items with few interactions to ensure the quality of the datasets \cite{NGCF,LightGCN,IMP-GCN}. Specifically, for all the datasets, we use the 10-core settings, which ensure that each user and item have at least 10 interactions. The detailed statistics of the three datasets are shown in Table~\ref{datasets-table}. 

We randomly split each dataset into training, validation, and testing set with a ratio of 80:10:10 for each user. For each observed user-item interaction, we treat it as a positive instance. Then, we randomly sample one negative item that the user did not consume before as a negative instance to pair the positive instance.

% Performance comparison with LightGCN at different layers
\begin{table*}
\renewcommand\arraystretch{1.4}
\centering
\caption{Performance comparison with LightGCN at different layers. The percentage in the brackets denote the relative performance improvement over LightGCN.}
\label{comparison_layers_with_lightgcn_table}
\resizebox{0.99\textwidth}{!}{
\begin{tabular}{c|c|ll|ll|ll}
\hline
\multicolumn{2}{c|}{Dataset} & \multicolumn{2}{c|}{Yelp2018} & 
\multicolumn{2}{c|}{Amazon-Book} & \multicolumn{2}{c}{Kindle-Store} \\
\hline
\#Layer & Method & Recall & NDCG & Recall & NDCG & Recall & NDCG \\
\hline \hline
\multirow{2}{*}{1 Layer} & 
LightGCN & 0.0631          & 0.0515          & 0.0384           & 0.0298           & 0.0964           & 0.0600 \\ &
QGCN     & 0.0633(+0.32\%) & 0.0519(+0.78\%) & 0.0489(+27.34\%) & 0.0376(+26.17\%) & 0.1250(+29.67\%) & 0.0788(+31.33\%) \\
\hline
\multirow{2}{*}{2 Layer} & 
LightGCN & 0.0622          & 0.0504          & 0.0411           & 0.0315           & 0.1021           & 0.0631 \\ &
QGCN     & 0.0656(+5.47\%) & 0.0538(+6.75\%) & 0.0480(+16.79\%) & 0.0364(+15.56\%) & 0.1244(+21.84\%) & 0.0779(+23.45\%) \\
\hline
\multirow{2}{*}{3 Layer} & 
LightGCN & 0.0639          & 0.0525          & 0.0410           & 0.0318           & 0.1040           & 0.0639 \\ &
QGCN     & 0.0662(+3.60\%) & 0.0546(+4.00\%) & 0.0464(+13.17\%) & 0.0353(+11.01\%) & 0.1205(+15.87\%) & 0.0749(+17.21\%) \\
\hline
\multirow{2}{*}{4 Layer} & 
LightGCN & 0.0649          & 0.0530          & 0.0406           & 0.0313          & 0.1024           & 0.0627 \\ &
QGCN     & 0.0668(+2.93\%) & 0.0547(+3.21\%) & 0.0448(+10.34\%) & 0.0340(+8.63\%) & 0.1167(+13.96\%) & 0.0725(+15.63\%) \\
\hline
\end{tabular}
}
\end{table*}

%****training loss and testing recall figure****
\begin{figure*}[t]
    \subfloat{\includegraphics[width=0.245\textwidth]{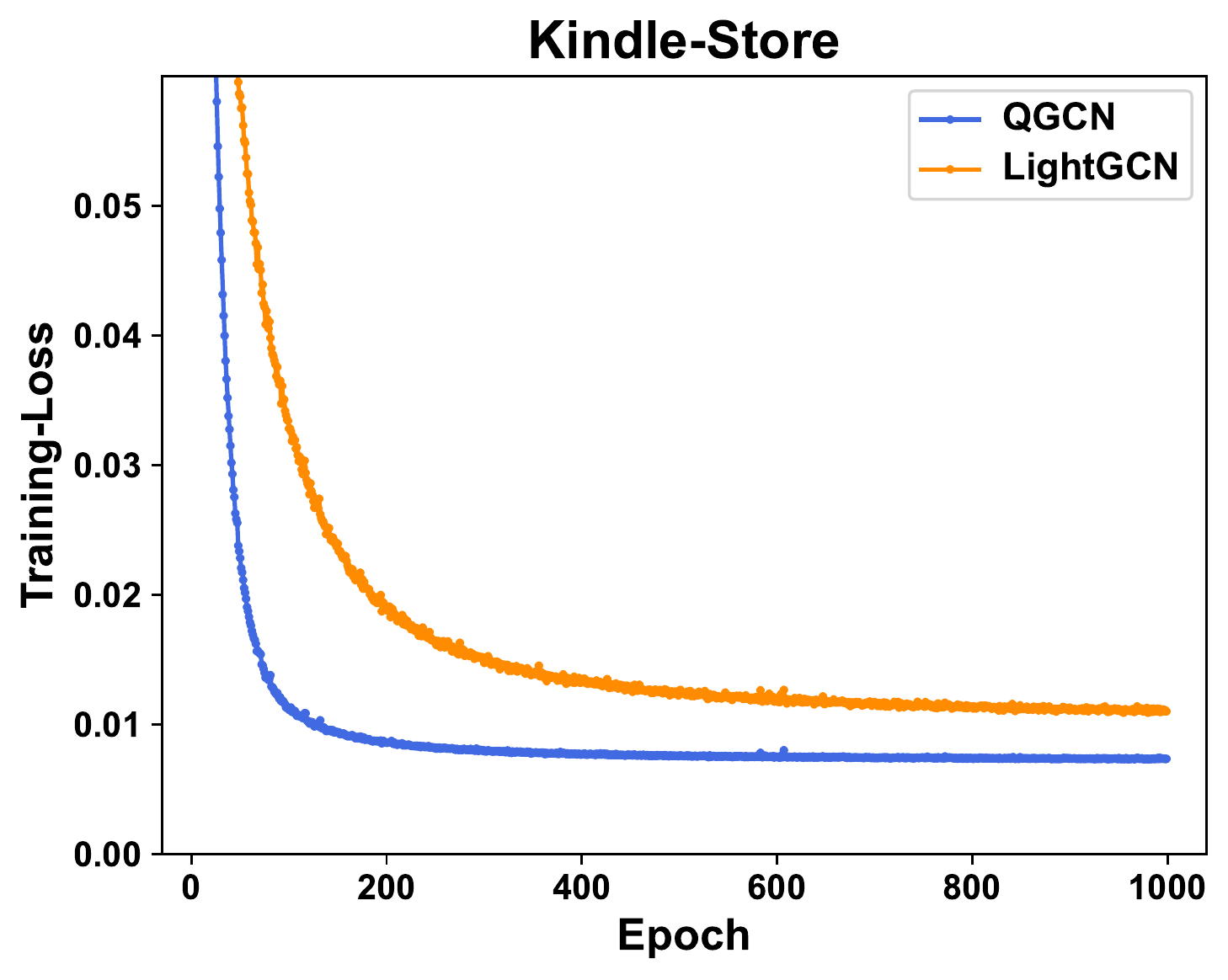}}
    \hfill
    \subfloat{\includegraphics[width=0.245\textwidth]{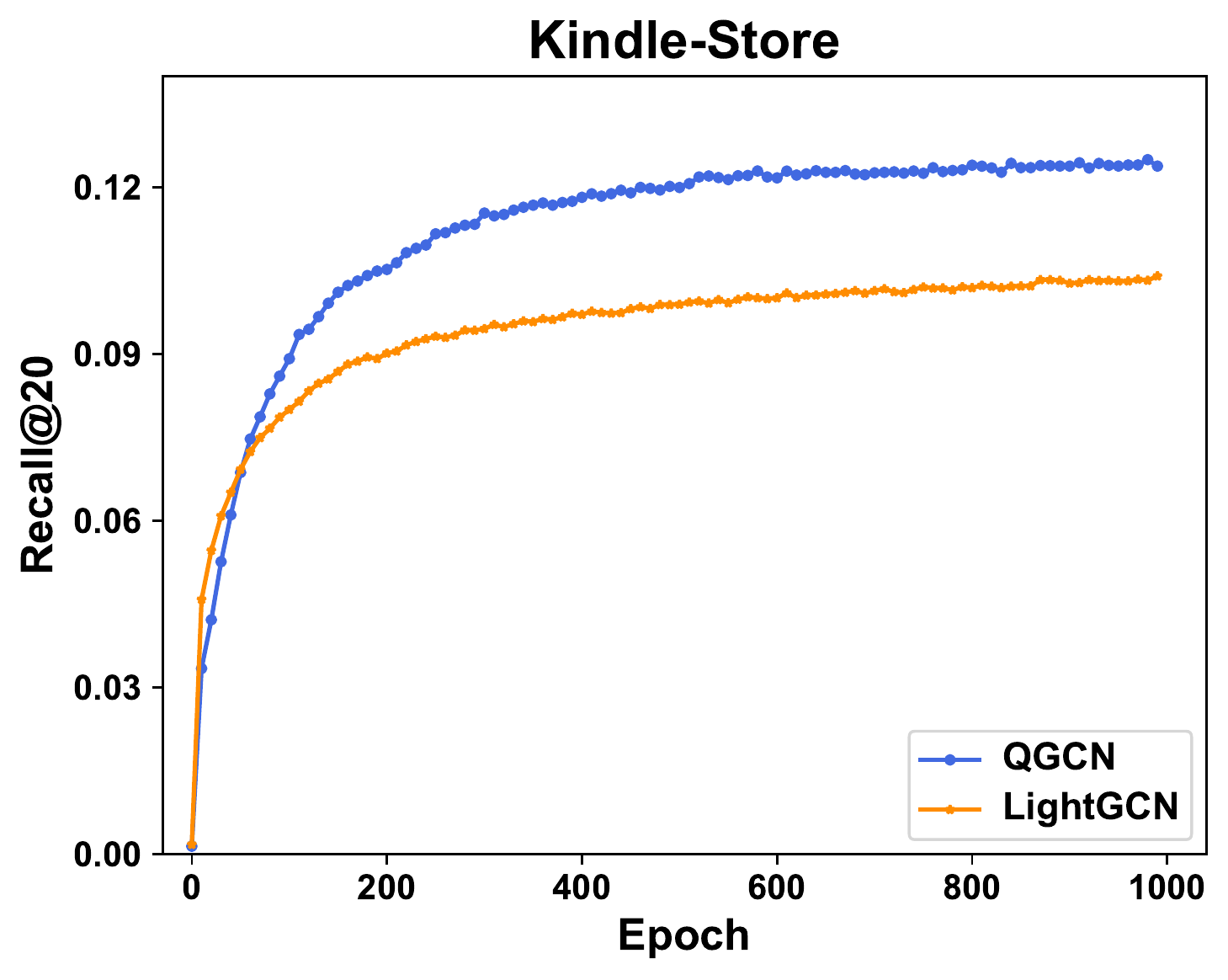}}
    \hfill
    \subfloat{\includegraphics[width=0.245\textwidth]{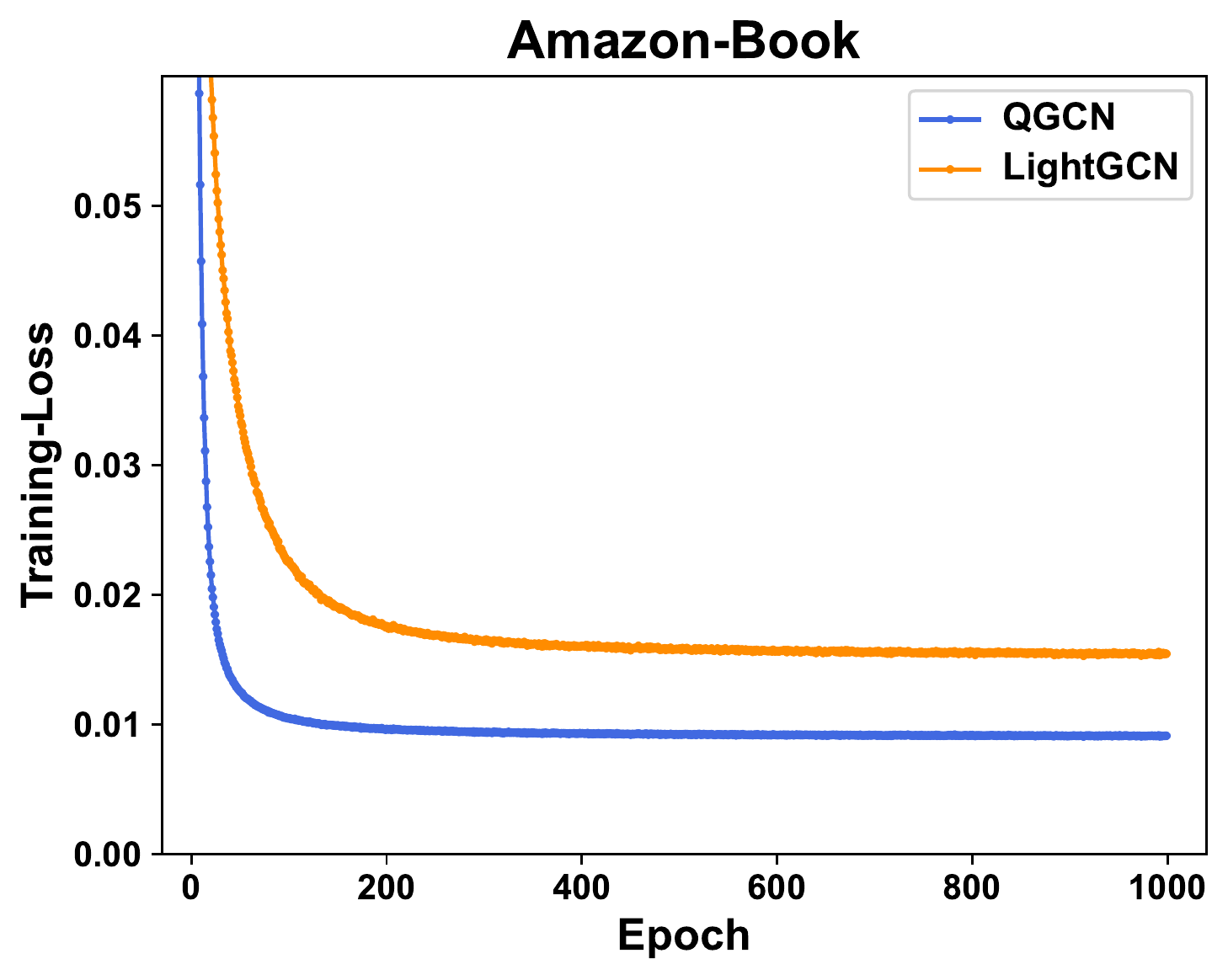}}
    \hfill 	
    \subfloat{\includegraphics[width=0.245\textwidth]{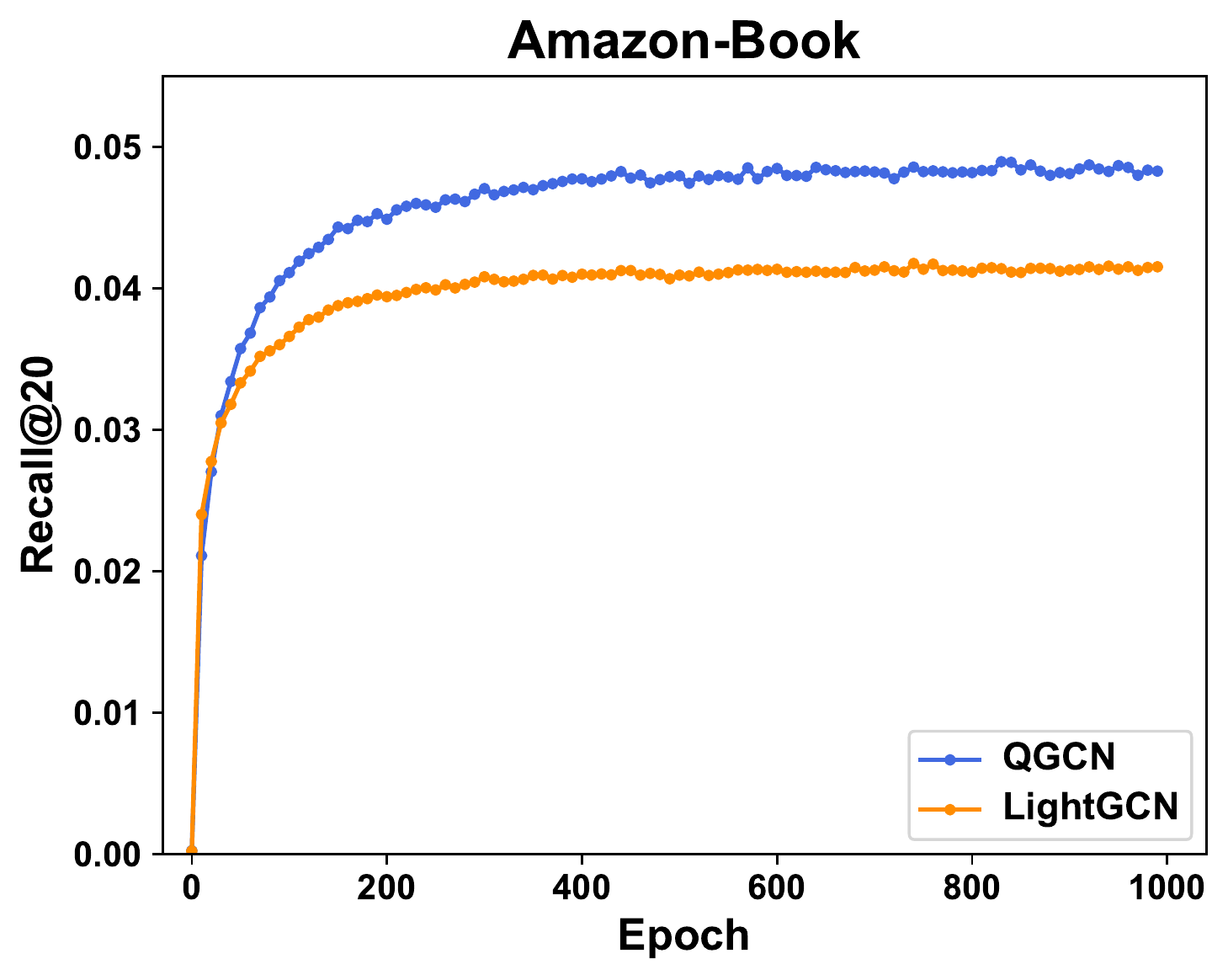}}
\caption{Training curves of QGCN and LightGCN, which are evaluated by training loss and testing recall per 10 epochs on Kindle-Store and Amazon-Book(results on Yelp2018 show the same trend which are omitted for space).} 
\label{training loss and testing recall curve}
\end{figure*}

\subsection{Experimental Settings}
\subsubsection{Evaluation Metrics}
To evaluate the effectiveness of our model on top-K recommendation, we take two evaluation metrics widely used in previous work: Recall@K and NDCG@K. Here, we set $K = 20$ by default, and the average results for all users in the testing set are reported. The specific definition is as follows:
\begin{itemize}
    \item [$\bullet $]  Recall@K describes the percentage of user-item rating records included in the final recommendation list. We denote the recommendation list for a user as $R_K$, and the corresponding testing set as $T$. Then, the specific definition of Recall@K is as follows:
    \begin{equation}
    \text{Recall@K} = \frac{\lvert T\cap R_K \rvert}{\lvert T \rvert}.
    \end{equation}
    \item [$\bullet $] NDCG@K \emph{i.e.} Normalized Discounted Cumulative Gain measures the quality of ranking, which emphasizes more on the relevance of the items on the top of the recommendation list. We denote the relevance of the $i$-th item in the recommendation list as $r_i$, and the set of relevant items as $R$. Then, the specific definition of NDCG@K is:
    \begin{equation}
        \text{NDCG@K} = \frac{\text{DCG@K}}{\text{IDCG@K}},
        % \text{NDCG@K} = \frac{\sum \limits _{i=1} ^ K \frac{r_i}{\log_2(i+1)}}{\sum \limits _{i=1} ^ {|R|} \frac{1}{\log_2(i+1)}},
        % \begin{split}
        %     \text{DCG@K} &= \sum \limits _{i=1} ^ K \frac{r_i}{\log_2(i+1)}, \\
        %     \text{IDCG@K} &= \sum \limits _{i=1} ^ {|R|} \frac{1}{\log_2(i+1)}, \\
        %     \text{NDCG@K} &= \frac{\text{DCG@K}}{\text{IDCG@K}},
        % \end{split}
    \end{equation}
    where DCG@K and IDCG@K are defined as follows:
    \begin{equation}
        \begin{split}
            \text{DCG@K} &= \sum \limits _{i=1} ^ K \frac{r_i}{\log_2(i+1)}, \\
            \text{IDCG@K} &= \sum \limits _{i=1} ^ {|R|} \frac{1}{\log_2(i+1)}. \\
        \end{split}
    \end{equation}
    % where $r_i$ denotes the relevance of the $i$-th item in the recommendation list, and $R$ denotes the set of relevant items.
\end{itemize}

\subsubsection{Baselines}
To demonstrate the effectiveness of our proposed QGCN model, we compare QGCN with the following competitive baseline methods:
\begin{itemize}
    \item [$\bullet $]{\textbf{NeuMF}} \cite{CF_based_model_2(NeuMF)}: NeuMF, a state-of-the-art neural collaborative filtering model, captures the non-linear interactions between user and item embeddings with multiple hidden layers.
    
    \item [$\bullet $]{\textbf{HOP-Rec}} \cite{HOP-Rec}: HOP-Rec, a state-of-the-art graph-based model, exploits the high-order connectivity between users and items by performing random walks to augment a user's interactions.
    
    \item [$\bullet $]{\textbf{GC-MC}} \cite{GC-MC}: GC-MC explores the first-order connectivity between users and items by utilizing only one convolution layer over the user-item bipartite graph.
    
    \item [$\bullet $]{\textbf{NGCF}} \cite{NGCF}: NGCF leverages the message-passing mechanism to obtain high-order connectivity and collaborative signal in the user-item integration graph.
    
    \item [$\bullet $]{\textbf{LightGCN}} \cite{LightGCN}: LightGCN removes two components, feature transformation and non-linear activation in NGCF, leading to improvement on training efficiency and generation ability.
\end{itemize}

\subsubsection{Parameter Settings}
We implement our QGCN model in PyTorch \footnote{https://pytorch.org}. The embedding size is fixed to 64 for all models. We optimize QGCN with Adam \cite{Adam-optimization} with the default learning rate of 0.0001 and set batch size as 2048 for speed. We apply a grid search for the only two hyper-parameters: the dropout rate is tuned among \{0.0, 0.1, 0.2, 0.3, 0.4, 0.5, 0.6, 0.7, 0.8\} and the coefficient of $L_2$ normalization in Equation~\ref{loss function} is searched in $\{1e^{-6}, 1e^{-5}, \dots, 1e^{-2}\}$.The embedding parameters are initialized with the Xavier method \cite{Xavier-initialization}.

\subsection{Performance Comparison (RQ1)}
\subsubsection{Comparison with LightGCN}
We conduct experiments under different graph convolution layer numbers for detailed comparison with LightGCN, and results are shown in Table~\ref{comparison_layers_with_lightgcn_table}. The percentage of relative improvement at each layer on Recall@20 and NDCG@20 is calculated as well. Specifically, the results of LightGCN on Yelp2018 and Amazon-Book are copied from its original paper \cite{LightGCN}, and we tune the hyper-parameters of LightGCN (\emph{i.e.} the $L_2$ regularization coefficient $\lambda$) on Kindle-Store and report with the optimal settings. We further plot the training curves of training loss and testing recall per 10 epochs on Kindle-Store and Amazon-Book with optimal settings on both LightGCN and our QGCN model in Fig.~\ref{training loss and testing recall curve}, where results on Yelp2018 show the same trend and are omitted for space. We summarize the main observations as follows:
\begin{itemize}
    \item [$\bullet $] In most cases, QGCN outperforms LightGCN by a large margin under different layer numbers ranging from 1 to 4 layers. The average improvement over all the datasets under different layer numbers is 19.11\%, 14.70\%, 10.88\% and 9.08\% \emph{w.r.t.} Recall@20 and 19.43\%, 15.25\%, 10.74\% and 9.16\% \emph{w.r.t.} NDCG@20. 
    The significant improvement under each layer number on each dataset indicates that the quaternion embedding and quaternion feature transformation enhance the representation learning a lot. 
    \item [$\bullet $] The improvements on Yelp2018 are relatively less significant, and the best results are gained at four layers. On Amazon-Book and Kindle-Store, our QGCN model gains huge and up to 30\%, relative performance improvement, and it can achieve the best results with only one quaternion embedding propagation layer. We ascribe this to the characteristics of the datasets, sparsity, and the excellent representation learning capability of the quaternion embedding and quaternion feature transformation. As we mentioned above, the quaternion feature transformation enhances the inter-latent interactions between real and imaginary components, enabling it to capture the graph structural features more effectively, distinguish the contribution of different nodes during message propagation, and thus improve both performance and robustness. Therefore, as the sparsity of the dataset decreases, the quaternion feature transformation could highlight its contribution to distilling sufficient information from the sparse user-item interaction graphs and further lead to more significant performance improvement. Moreover, the observations mentioned above that our QGCN model gaining huge relative performance improvement on sparse graphs is of great significance to the practical applications and real recommendation scenarios since real-world graphs are often extremely sparse.
    \item [$\bullet $] Our QGCN model obtains relatively lower training loss during the whole training process than that in LightGCN, which indicates our QGCN model can better fit the training data and further obtains better testing results. It demonstrates that with the quaternion embedding and quaternion feature transformation, our model obtains stronger generalization capability. 
\end{itemize}

% performance comparison with SOTA baselines
% For Single Column
\begin{table}[t]
\renewcommand\arraystretch{1.4}
\centering
\caption{Overall performance comparison over three datasets.}
\label{performance-comparison-table}
\resizebox{0.99\linewidth}{!}{
\begin{tabular}{c|cc|cc|cc}
\hline
Dataset & \multicolumn{2}{c|}{Yelp2018} & 
\multicolumn{2}{c|}{Amazon-Book} & \multicolumn{2}{c}{Kindle-Store} \\
\hline
Metric & Recall & NDCG & Recall & NDCG & Recall & NDCG \\
\hline \hline
NeuMF & 0.0451 & 0.0363 & 0.0258 & 0.0200 & 0.0496 & 0.0206 \\
\hline
HOP-Rec & 0.0517 & 0.0428 & 0.0309 & 0.0232 & 0.0796 & 0.0458\\
\hline 
GC-MC & 0.0462 & 0.0379 & 0.0288 & 0.0224 & 0.0793 & 0.0455 \\
NGCF & 0.0579 & 0.0477 & 0.0344 & 0.0263 & 0.0825 & 0.0509 \\
LightGCN & 0.0649 & 0.0530 & 0.0411 & 0.0315 & 0.1040 & 0.0639 \\
\hline \hline
QGCN & \textbf{0.0668} & \textbf{0.0547} & \textbf{0.0489} & \textbf{0.0376} & \textbf{0.1250} & \textbf{0.0788} \\
\%Improv. & 2.93\% & 3.21\% & 18.98\% & 19.37\% & 20.19\% & 23.32\% \\
\hline
\end{tabular}
}
\end{table}

\subsubsection{Comparison with SOTA Methods}
Table~\ref{performance-comparison-table} shows the performance with competing methods. The best results are highlighted in bold. From Table~\ref{performance-comparison-table}, we have the following observations:
\begin{itemize}
    \item [$\bullet $] NeuMF, a state-of-the-art neural collaborative filtering model, performs relatively poorly since it captures the connectivity between user and item embeddings in the embedding learning process rather than leveraging the high-order user-item interactions.
    
    \item [$\bullet $] Compared with NeuMF, GC-MC utilizes one convolution layer to explore the first-order connectivity between users and items and improve the performance, demonstrating the influence of first-order neighbors for representation learning. 
    
    \item [$\bullet $] HOP-Rec exploits the high-order connectivity between users and items by performing random walks to augment a user's interactions, resulting in better performance than GC-MC. NGCF performs much better over the above baselines. It leverages the message-passing mechanism to obtain high-order connectivity and collaborative signal in the user-item integration graph. LightGCN removes two components, feature transformation and non-linear activation in NGCF, leading to improvement in training efficiency and generation ability. 
    
    \item [$\bullet $] QGCN outperforms all the baselines by a large margin over all the datasets. In particular, compared with the strongest baseline, \emph{i.e} LightGCN, QGCN gains on average 14.03\% improvement \emph{w.r.t.} Recall@20 and 15.30\% improvement \emph{w.r.t.} NDCG@20 over all the datasets. The significant improvements reveal that QGCN can better capture high-order user-item connectivity and learn better user and item embeddings.
\end{itemize}

\subsection{Robustness Analysis (RQ2)}
\subsubsection{Random Edges Injection}
\begin{figure}[t]
    \subfloat{\includegraphics[width=0.5\linewidth]{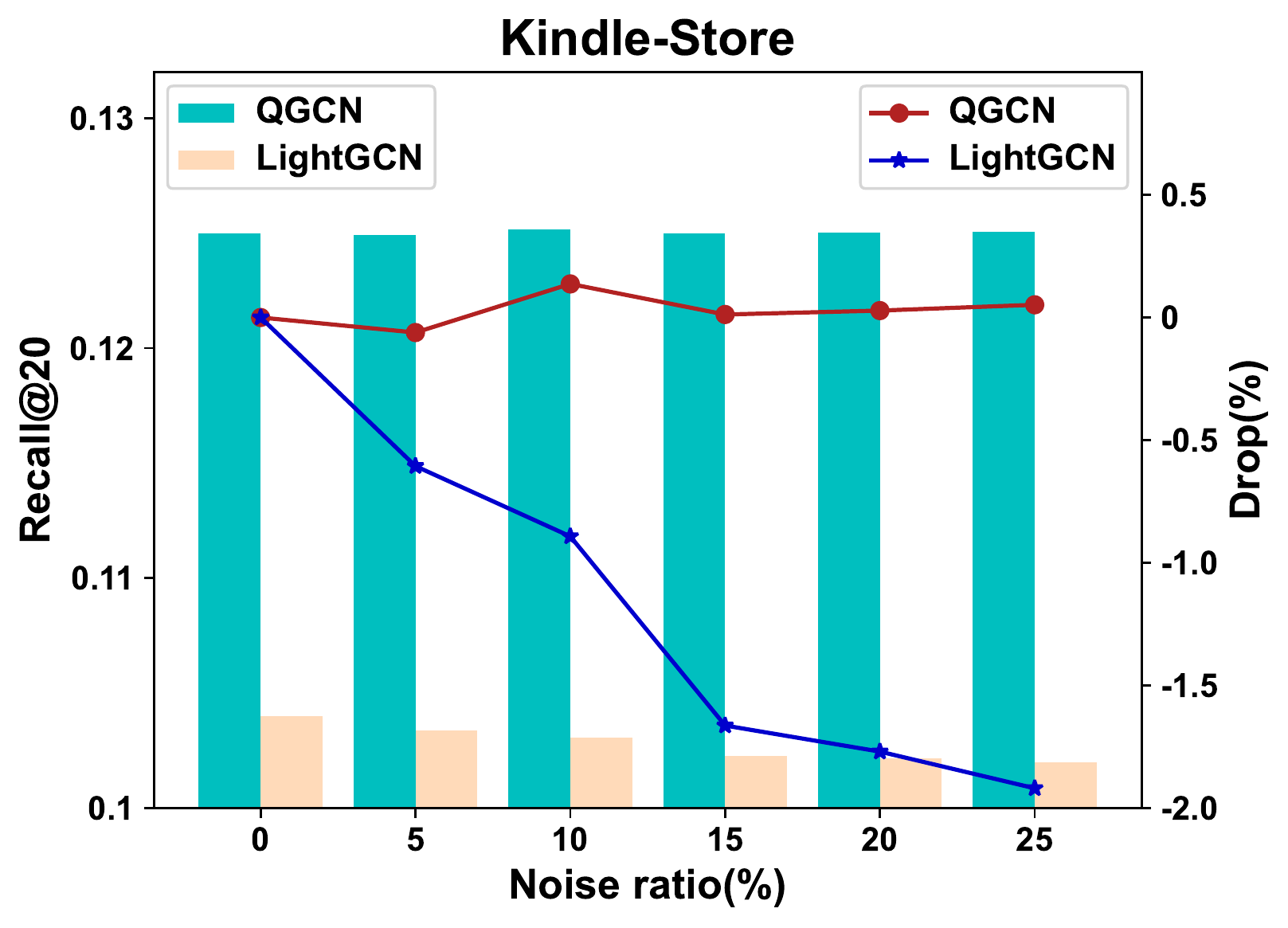}}
    \hfill
    \subfloat{\includegraphics[width=0.5\linewidth]{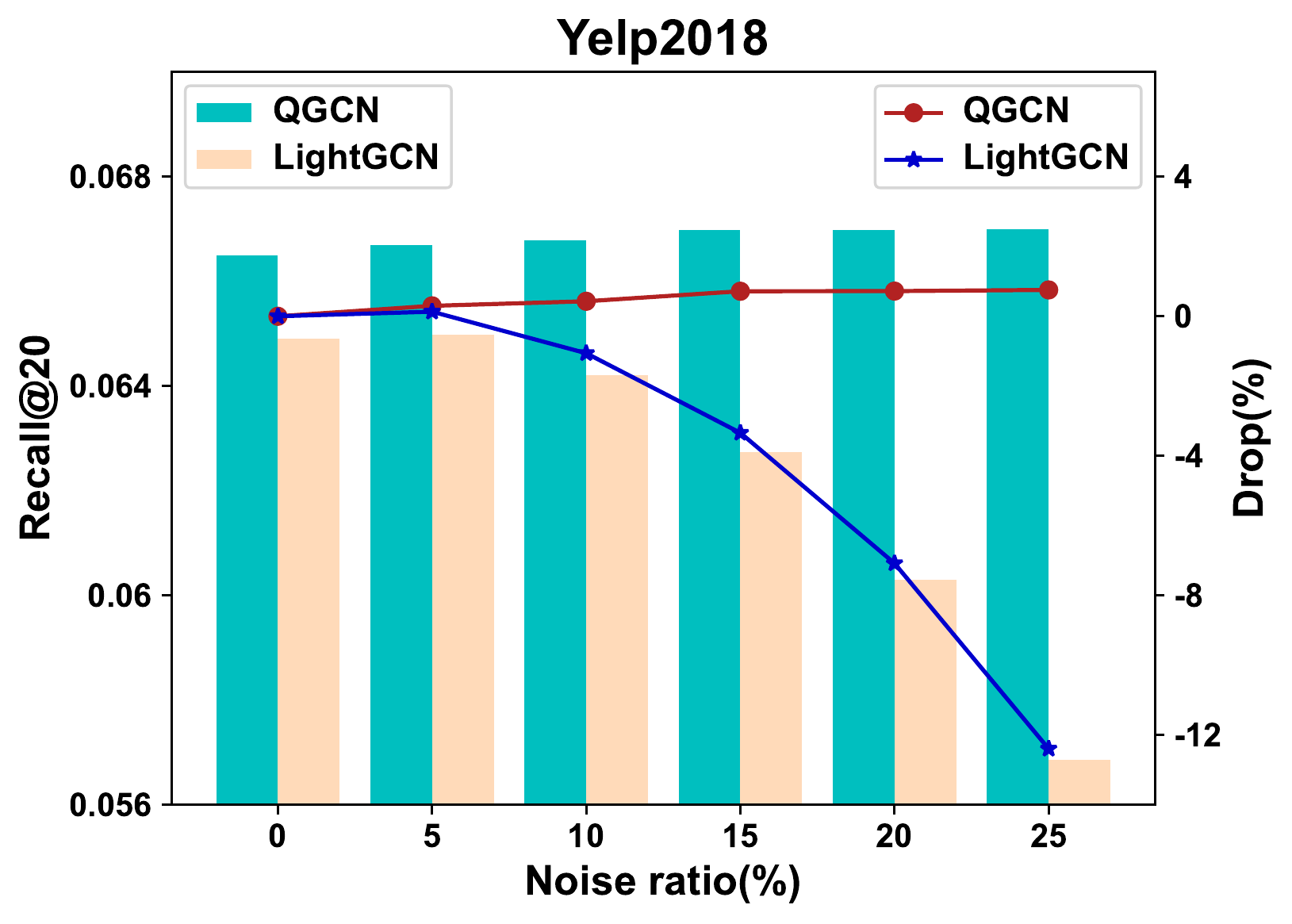}}
\caption{
Effect of random edges injection. The bar represents Recall@20, while the line represents the relative performance change compared to the original result. 
} 
\label{figure: adj_noise}
\end{figure}

To investigate the robustness of our QGCN model to noisy graphs, we conduct simulated experiments to explore the influence of random injection of edges. Specifically, we randomly connect the unobserved edges in the user-item interaction graph $\mathbf{R}$ as noisy edges to construct a noisy graph for the training process. The noise ratio is set in $\{5\%, 10\%, 15\%, 20\%, 25\% \}$. By the way, the compared LightGCN model and our QGCN model are trained with the same constructed noisy graph for a fair comparison. And we evaluate with the original graph (\emph{i.e.} 0\% edges injection). We further plot Recall@20 and relative drop compared with their original performance of both LightGCN and our QGCN model on Kindle-Store and Yelp2018 in Fig.~\ref{figure: adj_noise}.

%****figure: adj_noise****

%****figure: adj_dropedge****
\begin{figure}[t]
    \subfloat{\includegraphics[width=0.5\linewidth]{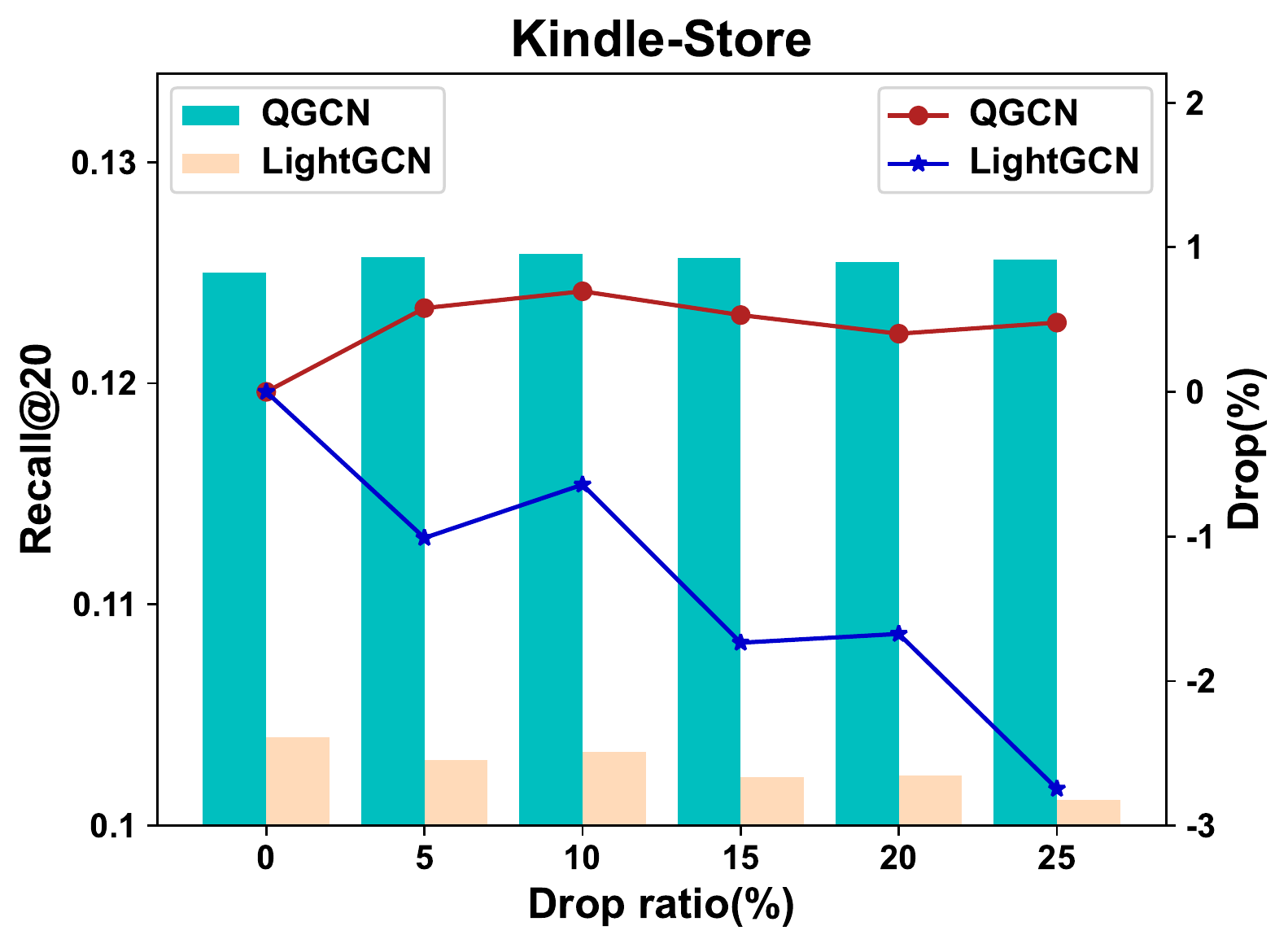}}
    \hfill
    \subfloat{\includegraphics[width=0.5\linewidth]{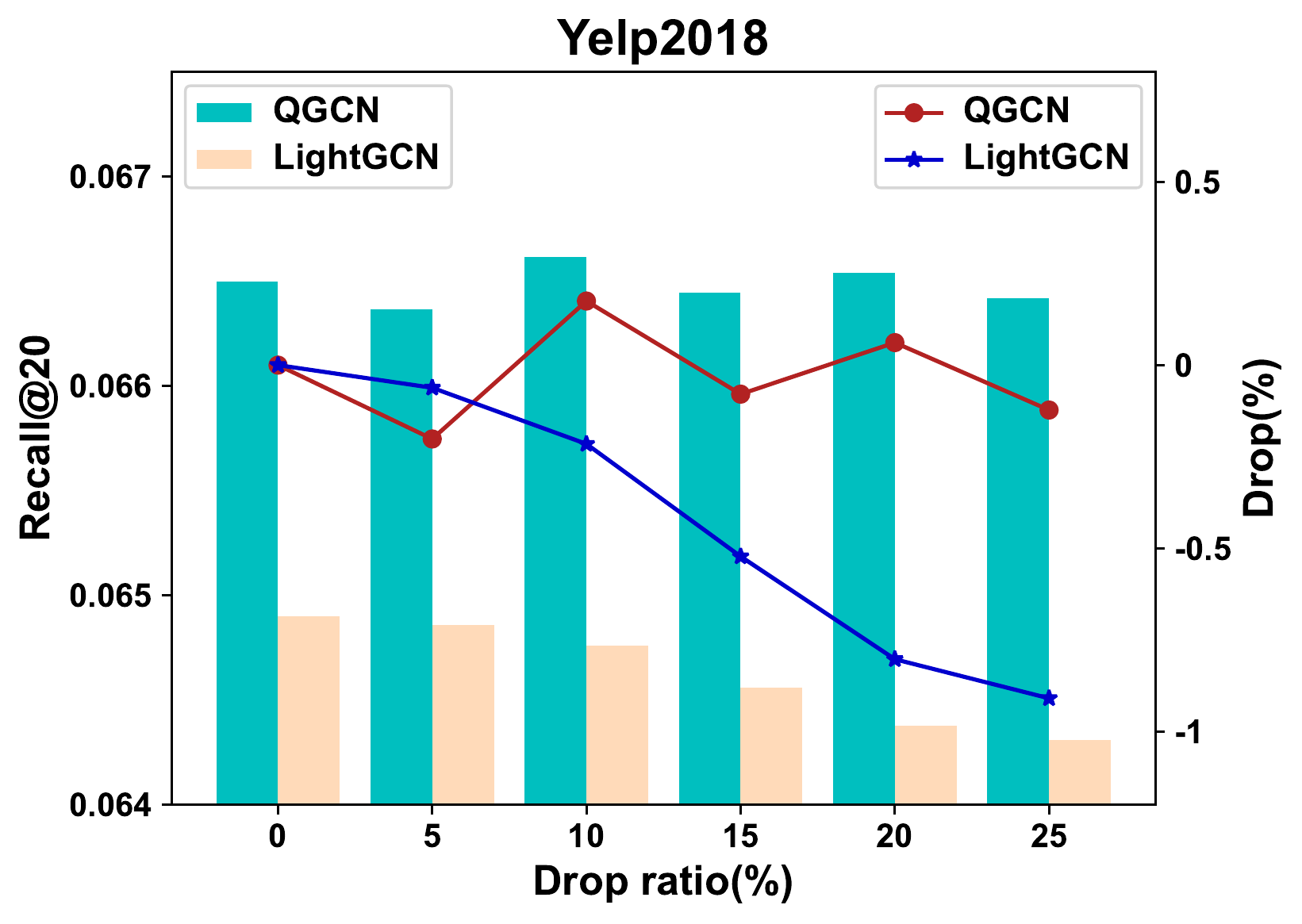}}
\caption{Effect of random edges discard. The bar represents Recall@20, while the line represents the relative performance change compared to the original result. } 
\label{figure: adj_dropedge}
\end{figure}

%****dropout rate effect figure****
\begin{figure*}[t]
    \subfloat{\includegraphics[width=0.33\textwidth]{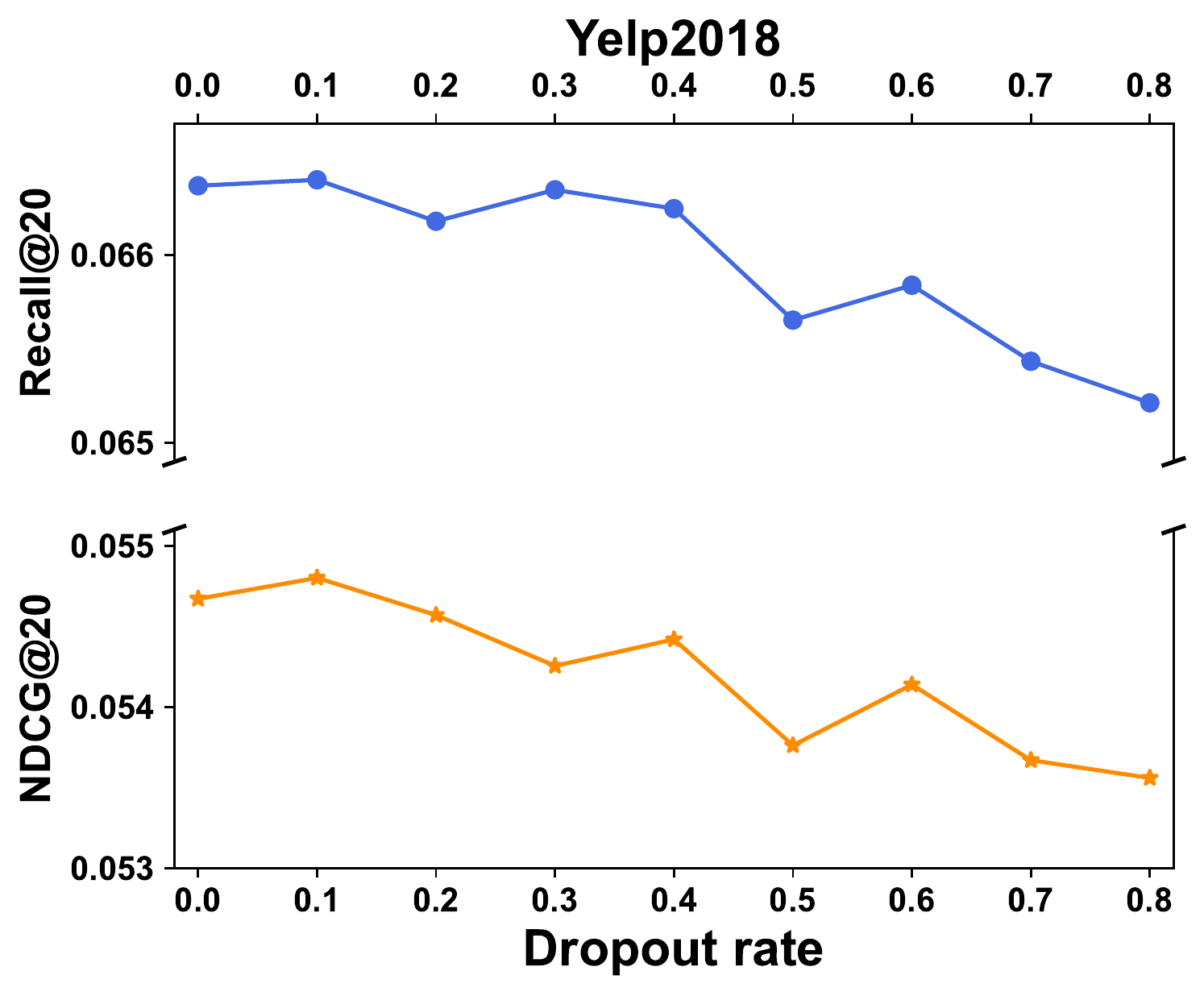}}
    \hfill
    \subfloat{\includegraphics[width=0.33\textwidth]{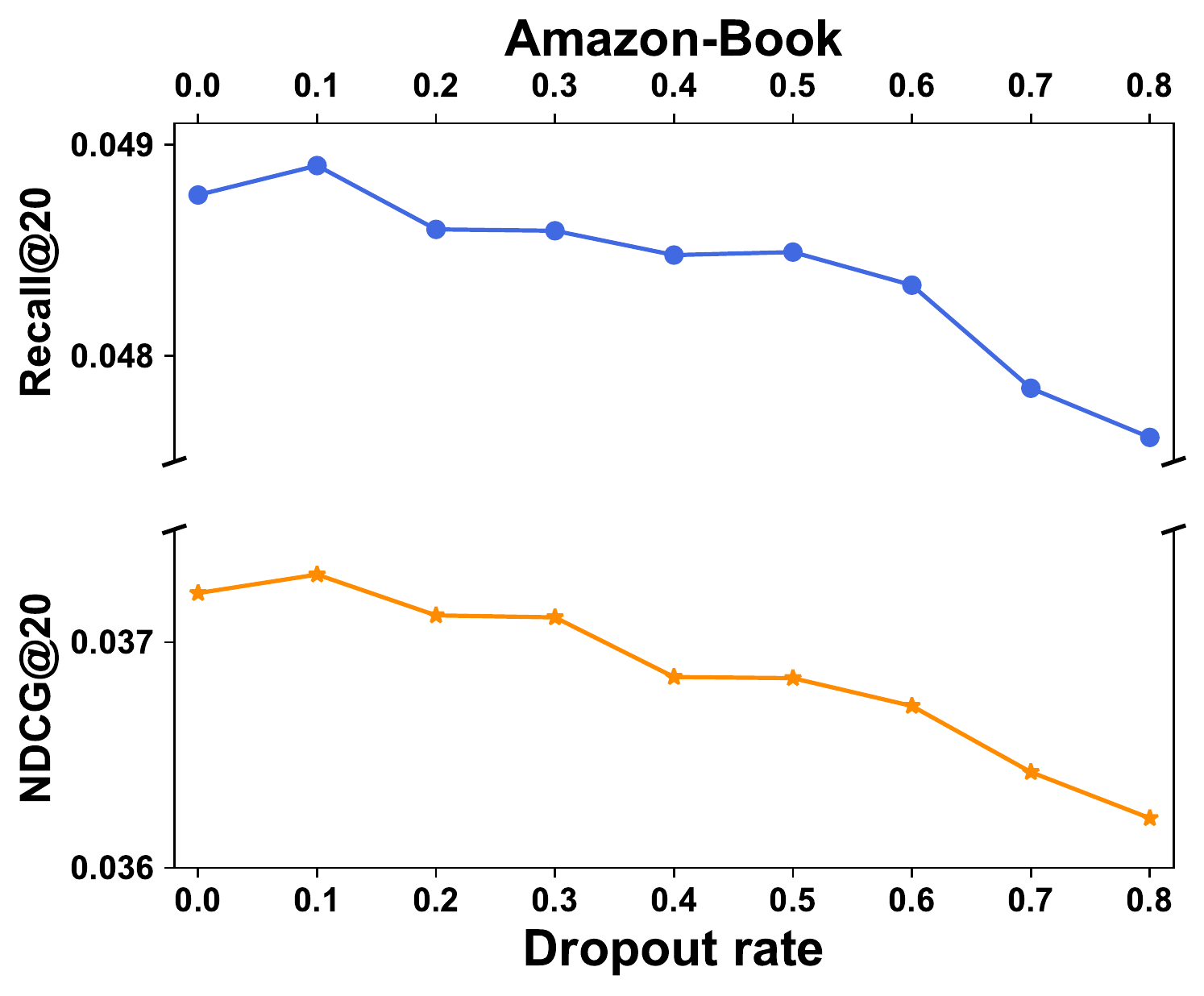}}
    \hfill
    \subfloat{\includegraphics[width=0.33\textwidth]{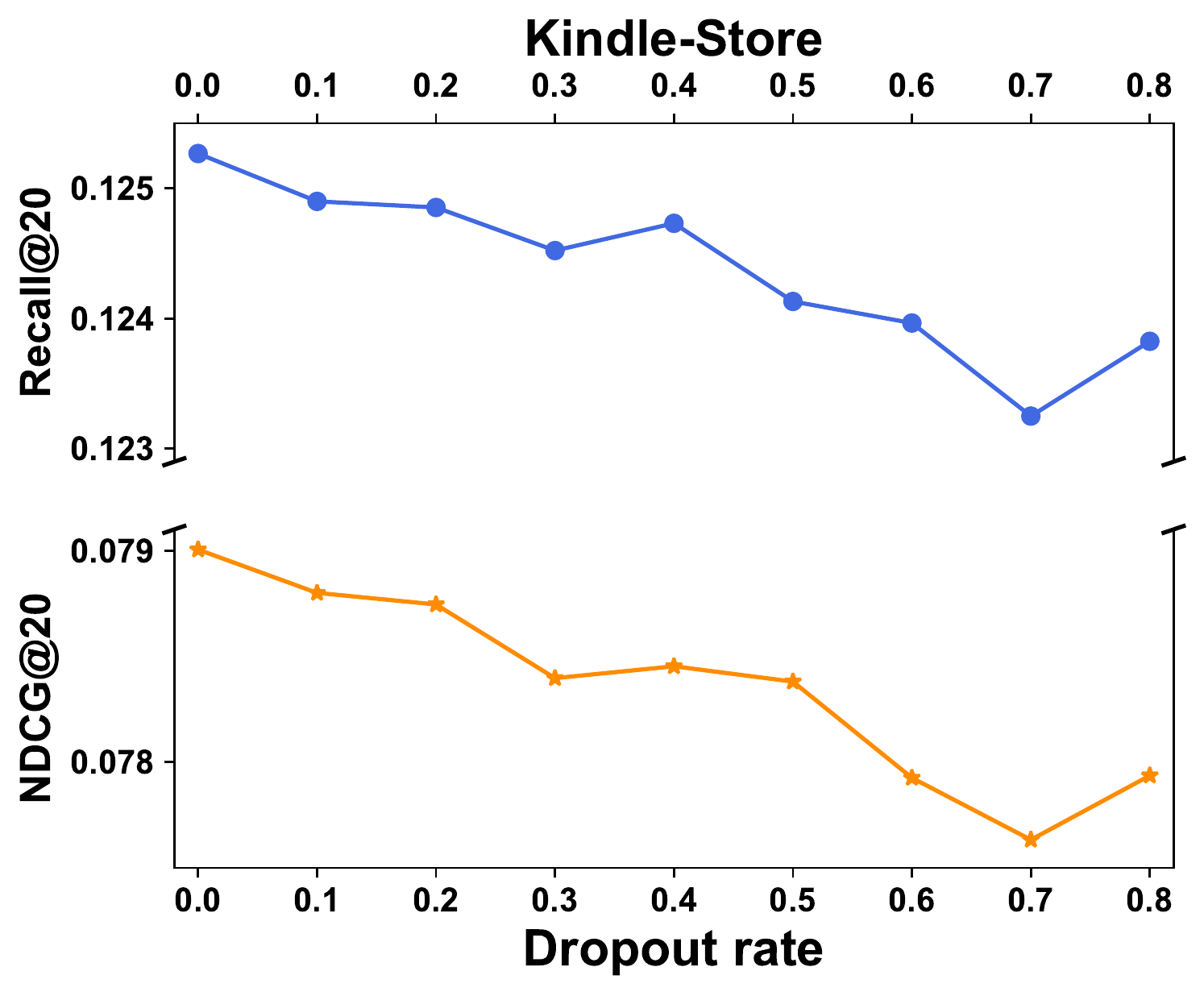}}
\caption{Effect of dropout rate.} 
\label{effect of Dropout rate figure}
\end{figure*}

We observe that QGCN consistently outperforms LightGCN by a large margin under different ratios of random edges injection on both Kindle-Store and Yelp2018. Along with the increase of the noise ratio, the performance of LightGCN decreases accordingly, while that of our QGCN model remains almost unchanged. For example, Recall@20 of LightGCN in the noisy graph with 25\% noise ratio of noise injection on Yelp2018 is 0.0568, dropping 12.48\% (\emph{i.e.} -12.48\%) compared to the original performance, 0.0649. In contrast to the large drop percent of LightGCN, the performance of our QGCN model under 25\% noise ratio even rises by 0.75\% (\emph{i.e.} +0.75\%) compared to that under 0\% noise ratio. The sharp decline of the relative drop of Recall@20 of LightGCN along with the increase of noise ratio reveals that LightGCN is extremely sensitive to noise, which is consistent with our argument mentioned before. Compared with the steep decline curve of Recall@20 of LightGCN, the relative performance change curve of our QGCN model is more steady, which demonstrates the robustness of our QGCN model to noisy graphs.

%****ablation study table****
\begin{table}[t]
\renewcommand\arraystretch{1.4}
\centering
\caption{Performance of our model and its variants.}
\label{ablation study table}
\resizebox{0.99\linewidth}{!}{
\begin{tabular}{c|cc|cc|cc}
\hline
Dataset & \multicolumn{2}{c|}{Yelp2018} & 
\multicolumn{2}{c|}{Amazon-Book} & \multicolumn{2}{c}{Kindle-Store} \\
\hline
Metric & Recall & NDCG & Recall & NDCG & Recall & NDCG \\
\hline \hline
QGCN-Q & 0.0603 & 0.0491 & 0.0369 & 0.0280 & 0.0939 & 0.0576 \\
QGCN-W & 0.0660 & 0.0541 & 0.0485 & 0.0370 & 0.1244 & \textbf{0.0795} \\
QGCN & \textbf{0.0668} & \textbf{0.0547} & \textbf{0.0489} & \textbf{0.0376} & \textbf{0.1250} & 0.0788 \\
\hline
\end{tabular}
}
\end{table}

\subsubsection{Random Edges Discard}
In addition to the characteristic of real-world user-item graphs containing a lot of noise, they are often incomplete as well. Thus, besides the simulated experiments on exploring the influence of random injection of edges, we also conduct experiments to explore the influence of the random discard of edges. Similarly, we construct a corrupted graph by randomly disconnect the existing edges in the user-item interaction graph $\mathbf{R}$ with a drop ratio ranging in $\{5\%, 10\%, 15\%, 20\%, 25\% \}$. We then train the compared LightGCN model and our QGCN model with the corrupted graph and evaluate with the original graph (\emph{i.e.} 0\% edges discard). The details of Recall@20 and relative drop are shown in Fig.~\ref{figure: adj_dropedge}.

We have similar observations from Fig.~\ref{figure: adj_dropedge}. Specifically, QGCN consistently outperforms LightGCN by a large margin \emph{w.r.t} different ratios of random edges discard on both Kindle-Store and Yelp2018. The steep performance decline curve of LightGCN is in sharp contrast to the steady curve of QGCN, demonstrating the robustness of our QGCN model to corrupted graphs. 

The simulated experiments on exploring the influence of random injection and discard of edges both demonstrate the robustness of our QGCN model. We ascribe this to the expressive quaternion feature transformation, distinguishing the contribution of different nodes and effectively capturing the graph structural features during message propagation. Thus, it can aggregate more useful information and further lead to better model performance and robustness. 

%****readout function table****
\begin{table}[t]
\renewcommand\arraystretch{1.4}
\centering
\caption{Influence of readout function.}
\label{readout function table}
\resizebox{0.99\linewidth}{!}{
\begin{tabular}{c|cc|cc|cc}
\hline
Dataset & \multicolumn{2}{c|}{Yelp2018} & 
\multicolumn{2}{c|}{Amazon-Book} & \multicolumn{2}{c}{Kindle-Store} \\
\hline
Metric & Recall & NDCG & Recall & NDCG & Recall & NDCG \\
\hline \hline
Max    & 0.0501 & 0.0387 & 0.0412 & 0.0312 & 0.1033 & 0.0655 \\
Sum    & 0.0429 & 0.0541 & 0.0467 & 0.0363 & 0.1206 & 0.0771 \\
Concat & 0.0572 & 0.0491 & 0.0475 & 0.0364 & 0.1222 & 0.0776 \\
Mean & \textbf{0.0668} & \textbf{0.0547} & \textbf{0.0489} & \textbf{0.0376} & \textbf{0.1250} & \textbf{0.0788} \\
\hline
\end{tabular}
}
\end{table}

\subsection{Ablation Study (RQ3)}
\subsubsection{Influence of Components}
We perform ablation studies to explore the contribution of different components to the model performance by comparing QGCN with the following two variants:
\begin{itemize}
    \item [$\bullet $]{\textbf{QGCN-Q}}: In this variant, we embed all users and items into the real-value space instead of the Quaternion space and maintain the component of feature transformation.
    \item [$\bullet $]{\textbf{QGCN-W}}: This variant removes the quaternion transformation matrices during message propagation.
\end{itemize}

Table~\ref{ablation study table} shows the results of the two variants of QGCN, and the best results are highlighted in bold. QGCN performs much better than QGCN-Q, which shows the significant influence of modeling in the Quaternion space. And QGCN outperforms QGCN-W in most cases, indicating the effectiveness of quaternion transformation matrices. The comparison between QGCN and its two variants demonstrates that the design of our proposed QGCN model is reasonable and effective.

\subsubsection{Influence of Readout Function}
Since different pooling methods generate different final user and item embeddings, we conduct experiments and investigate the influence of the readout function applied to our model. Table~\ref{readout function table} shows the results under different readout functions, and the best results are highlighted in bold. We can observe that Mean pooling performs relatively better than the other three readout functions, Max, Sum, Concat pooling. We think Mean pooling method could not only maintain the information of nodes but also uniform the user and items representations generated at each layer, leading to more powerful generalization capability.

\subsection{Hyper-parameter Study (RQ4)}
\subsubsection{Effect of Dropout Rate}
Dropout drops the units of the neural networks with a certain probability during the training process, which proves to be an effective way to prevent neural networks from overfitting \cite{dropout_origin_1, dropout_origin_2}. Motivated by the previous work of introducing dropout into graph convolutional network \cite{dropout_in_GCN} and GCN-based recommendation models \cite{NGCF}, we investigate the influence of the dropout rate $p$ ranging from 0.0 to 0.8 on our proposed QGCN model. 

Fig.~\ref{effect of Dropout rate figure} displays the experimental results, including Recall@20 and NDCG@20, under different dropout rates over all the datasets. For Yelp2018 and Amazon-Book, the dropout rate set as 0.1 leads to the best performance, while that set as 0.0 leads to the best performance on Kindle-Store. Besides, the performance degrades generally after the peak in that too many neurons lost leads to underfitting and limits the expression of our model. These observations are consistent with the findings of prior effort \cite{NGCF} and demonstrate the effectiveness of proper dropout rate settings in our model.

%****regularization effect figure****
\begin{figure}[t]
    \subfloat{\includegraphics[width=0.5\linewidth]{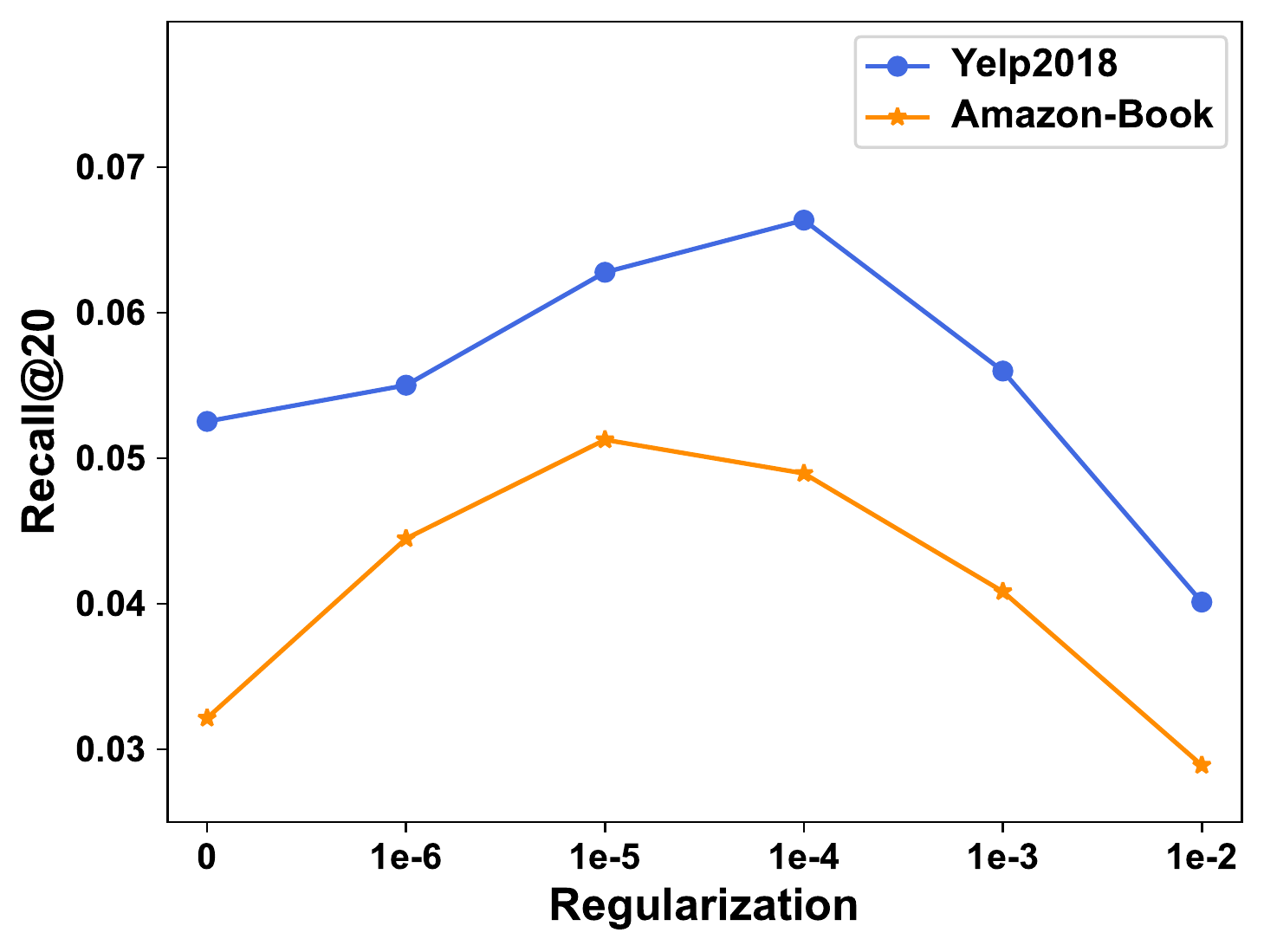}}
    \hfill
    \subfloat{\includegraphics[width=0.5\linewidth]{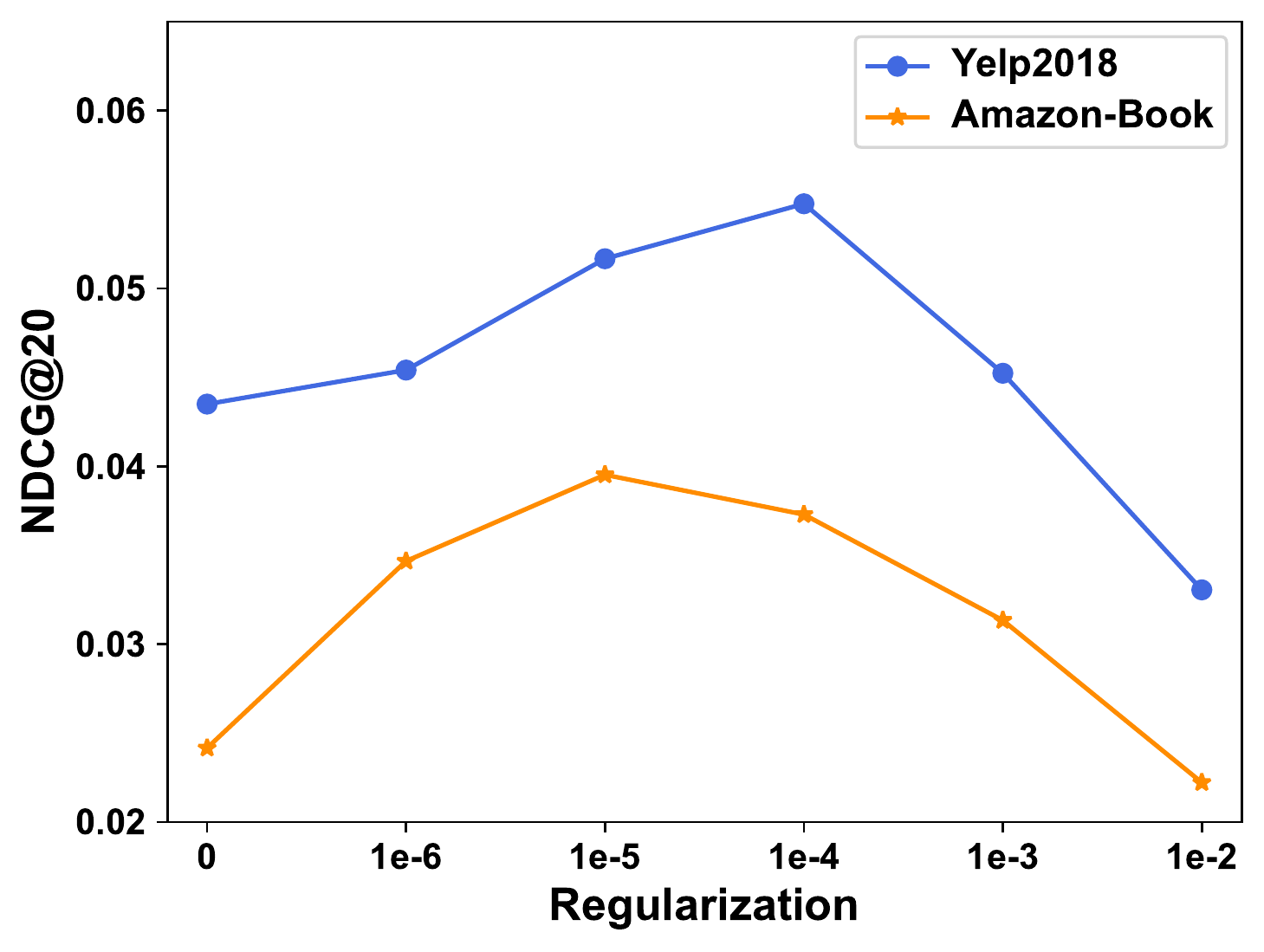}}
\caption{Effect of regularization.} 
\label{effect of regularization figure}
\end{figure}

\subsubsection{Effect of Regularization}
Regularization is an effective strategy to prevent overfitting, so that we tune the coefficient of $L_2$ normalization $\lambda$ among $\{1e^{-6}, 1e^{-5}, \dots, 1e^{-2}\}$ to investigate the influence of the regularization on our proposed model.

Fig.~\ref{effect of regularization figure} shows the performance of our QGCN model under different regularization coefficients $\lambda$ on Yelp2018 and Amazon-Book, and the effect of regularization over Kindle-Store are omitted for exactly the same trend. As shown in Fig.~\ref{effect of regularization figure}, too small or too large regularization coefficient result in relatively poor performance. Results are relatively steady when the regularization coefficient $\lambda$ is set between $1e^{-5}$ and $1e^{-4}$, while the performance significantly decrease when $\lambda$ is set larger than $1e^{-4}$ or smaller than $1e^{-5}$. This indicates that a medium regularization coefficient is more suitable for our model. Specifically, the optimal regularization coefficient for Yelp2018, Amazon-Book, and Kindle-Store is $1e^{-4}$, $1e^{-5}$ and $1e^{-4}$ respectively. 

\section{RELATED WORK}
\subsection{Quaternion-Based Applications}
Quaternion space is a hyper-complex vector space, where each quaternion is a hyper-complex number consisting of one real and three imaginary components. Owing to Hamilton product, which is the multiplication of quaternions, the interactions between real and imaginary components of two quaternions are enhanced, leading to highly expressive computations and up to four times reduction of parameters. In addition, if any slight change happens in the input quaternion, Hamilton product will generate an entirely different output \cite{Quaternion_expressive} and further influence the final performance. The Quaternion space has been successfully employed in various fields. 
For example, \cite{Q_signal_processing} applies a quaternionic Fourier transform and a quaternionic Gabor filter and exploits the symmetries inherent in the quaternion to find differences between subtly varying images.  
\cite{Q_image_classification_1} explores the benefits of generalizing one step further into the Quaternion space and provides the architecture components needed to build deep quaternion networks. 
\cite{Q_image_classification_2} re-designs the basic modules like convolution layer and fully-connected layer in the quaternion domain, which can be used to establish fully-quaternion convolutional neural networks, and results show that they outperform the real-valued CNNs with the same structures. 
\cite{Q_speech_recognition_2} applies the Quaternion space into recurrent neural network (RNN) and long-short term memory neural network (LSTM) and achieves better performance than the basic model in a realistic application of automatic speech recognition. 
\cite{Q_speech_recognition_3} integrates multiple feature views in quaternion-valued convolutional neural network (QCNN), to be used for sequence-to-sequence mapping with the CTC model. 
\cite{Q_speech_recognition_1} investigates modern quaternion-valued models such as convolutional and recurrent quaternion neural networks in the context of speech recognition.

Recently, there has been some work introducing the Quaternion space into graph representation learning to obtain more expressive graph-level representation \cite{QGNN,QuatRE,QKGE}. For example, \cite{QGNN} generalizes graph neural networks within the Quaternion space for graph classification, node classification, and text classification. \cite{QuatRE,QKGE} introduce more expressive quaternion representations to model entities and relations for knowledge graph embeddings for knowledge graph completion. 

\subsection{Collaborative Filtering}
Collaborative Filtering (CF) based models \cite{CF_based_model_1,CF_based_model_2(NeuMF),CF_based_model_3,CF_based_model_4,CF_based_model_5} have shown great performance in learning user and item representations. Matrix factorization \cite{MF_for_Rec} and Neural collaborative filtering model \cite{CF_based_model_2(NeuMF)} are widely used CF models, which embed users and items into the latent space. 
Some methods consider a user's historical interactions as his or her feature, such as FISM \cite{FISM} and SVD++ \cite{SVD++} which lead to a better user representation. In addition, side information is leveraged to further improve the recommendation quality like image \cite{image_enhanced_CF(VBPR)}, review \cite{review_enhanced_CF} and knowledge graph \cite{KG_enhanced_CF(CKAN),KG_enhanced_CF(KGAT),KG_enhanced_CF(RippleNet),KG_enhanced_CF(CollaborativeKowledgeBase)}. 
Recently, attention mechanisms have been widely introduced to recommendation models, such as ACF \cite{ACF} and NAIS \cite{NAIS}, in order to capture the different contributions of a user's historical interactions, improving their performance a lot.

\subsection{Graph-Based Recommendation}
Another research line exploits the user-item interaction graph for recommendation. Prior efforts like ItemRank \cite{ItemRank}, adopt label propagation on the graph and encourage connected nodes to have similar labels. HOP-Rec \cite{HOP-Rec} firstly performs random walks to augment a user's interactions. The powerful performance of HOP-Rec over MF suggests that exploiting the connectivity information leads to better user and item representations. However, HOP-Rec relies on random walks and is unable to explore the high-order connectivity between users and items, leading to careful tuning efforts indispensable.

Recently, GCN-based recommendation models have surged to learn better user and item representations in user-item bipartite graphs. For example, PinSage \cite{PinSage} combines random walk and graph convolutions to learn the embeddings of nodes. GC-MC explores the first-order connectivity between users and items by utilizing only one convolution layer over the user-item bipartite graph. NGCF \cite{NGCF} leverages the message-passing mechanism to obtain high-order connectivity and collaborative signal in the user-item integration graph. LightGCN \cite{LightGCN} removes two components, feature transformation and non-linear activation in NGCF \cite{NGCF}, leading to improvement in training efficiency and generation ability. 

We move a step further on this research line. Despite the great success of existing GCN-based recommendation models, GCN is still vulnerable to noisy and incomplete graphs, which are common in real-world scenarios, due to its recursive message propagation mechanism \cite{SGCN,vulnerable_to_the_quality_of_graph_1,vulnerable_to_the_quality_of_graph_2}. However, some latest GCN-based recommendation models ($e.g.$ LightGCN \cite{LightGCN}) remove the feature transformation during message propagation, making it unable to effectively capture the graph structural features and become more sensitive to noisy or missing information. Moreover, they model users and items in the Euclidean space, which has been demonstrated to have high distortion when modeling complex graphs \cite{HGCN,HGNN}, further degrading the capability to capture the graph structural features and leading to sub-optimal performance. Therefore, we move beyond the Euclidean space and fully utilize the Quaternion space, a hyper-complex space, to learn better user and item representations and feature transformation and thus improve both performance and robustness. 

\section{CONCLUSION}
In this work, we argued the limitation of the unreasonable operation of removing the feature transformation and modeling users and items in the Euclidean space and performed empirical studies to justify this argument. 
We moved beyond the Euclidean space, fully utilized the Quaternion space, a hyper-complex space, and proposed a simple yet effective Quaternion-based Graph Convolution Network model formed by a Quaternion Embedding Layer, Quaternion Embedding Propagation Layers, and a Prediction Layer.
Specifically, we first embedded all users and items into the Quaternion space with quaternion embeddings. Then, we introduced the quaternion embedding propagation layers with quaternion feature transformation to perform message propagation for aggregating more useful information. Finally, we combined the embeddings generated at each layer with the mean pooling strategy to obtain the final embeddings for recommendation. 
Extensive experiments on three public benchmark datasets were conducted to evaluate the effectiveness of our proposed model. Results showed that our model outperforms the state-of-the-art methods by a large margin. This indicates that it can better learn user and item representations. Besides, further robustness analysis demonstrated that our QGCN model is more robust to noisy and incomplete graphs and can effectively capture the graph structural features. Moreover, specific performance comparison showed that our QGCN model gains huge performance improvement on sparse graphs, which is of great significance to the practical applications and real recommendation scenarios. 

This work represents an attempt to explore the Quaternion space to model users and items and the effectiveness of quaternion transformation in the Quaternion-based GCN collaborative filtering methods. We believe the insights in this study are enlightening for introducing the Quaternion space into other recommendation scenarios and digging into the nature and effectiveness of quaternion transformation. 

\section*{ACKNOWLEDGMENTS}
This research was partially supported by NSFC (No. 61876117, 61876217, 61872258, 61728205), Exploratory Self-selected Project of the State Key Laboratory of Software Development Environment, and Priority Academic Program Development of Jiangsu Higher Education Institutions.

\bibliographystyle{IEEEtran}
\bibliography{reference}

\end{document}